\documentclass{template/egpubl}
\JournalPaper

\usepackage[T1]{fontenc}
\usepackage{template/dfadobe}
\biberVersion
\BibtexOrBiblatex
\usepackage[backend=biber,bibstyle=template/EG,citestyle=alphabetic,backref=true]{biblatex}
\electronicVersion
\PrintedOrElectronic
\ifpdf
    \usepackage[pdftex]{graphicx} \pdfcompresslevel=9
\else
    \usepackage[dvips]{graphicx} \fi
\usepackage{template/egweblnk} 

\usepackage[table]{xcolor}                  % for \definecolor, \rowcolors
\usepackage{amssymb}                        % for \mathbb, \mathfrak
\usepackage{xspace}                         % for \xspace
\usepackage{pgfplots}                       % for tikzpicture environment
\usepackage{hyperref}                       % for \href
\usepackage[ranges=true]{custom/acronyms}   % acronyms support
\usepackage{custom/orcid}
\usepackage{custom/nicematrix}

\newcommand{\eg}{e.g.\xspace}           % e.g.
\newcommand{\ie}{i.e.\xspace}           % i.e.
\newcommand{\etal}{et~al.\xspace}       % et al.
\newcommand{\enquote}[1]{``#1''}

\newcommand{\nD}[1]{#1D\xspace}         % 1D, 2D, 3D, ...
\newcommand{\D}{\nD{1}}                 % 1D
\newcommand{\DD}{\nD{2}}                % 2D
\newcommand{\DDD}{\nD{3}}               % 3D

\newcommand{\SO}[1]{\mathrm{SO}(#1)}    % SO(n)
\newcommand{\SE}[1]{\mathrm{SE}(#1)}    % SE(n)
\newcommand{\so}[1]{\mathfrak{so}(#1)}  % so(n)
\newcommand{\se}[1]{\mathfrak{se}(#1)}  % se(n)
\newcommand{\R}[1]{\mathbb{R}^{#1}}     % R^n
\newcommand{\header}[1]{%
    \normalsize \textbf{#1}%
}

\newcommand{\subheader}[1]{%
    \scriptsize #1%
}

\newcommand{\newurl}[2]{%
    \expandafter\newcommand\csname #1\endcsname{\href{#2}{</>}}%
}
\newcommand{\useurl}[1]{%
    \ifcsname #1\endcsname \csname #1\endcsname \fi%
}

\newurl{data:CM03}{http://mocap.cs.cmu.edu}
\newurl{data:IP14}{http://vision.imar.ro/human3.6m}
\newurl{data:HS16}{https://theorangeduck.com/page/deep-learning-framework-character-motion-synthesis-and-editing}
\newurl{data:SL16}{https://rose1.ntu.edu.sg/dataset/actionRecognition}
\newurl{data:LS20}{https://rose1.ntu.edu.sg/dataset/actionRecognition}
\newurl{data:VH18}{https://virtualhumans.mpi-inf.mpg.de/3DPW}
\newurl{data:MR07}{http://resources.mpi-inf.mpg.de/HDM05}
\newurl{data:MI21}{https://www.mixamo.com/}
\newurl{data:MG19}{https://amass.is.tue.mpg.de}

\newurl{code:JZ16}{https://www.github.com/asheshjain399/RNNexp}
\newurl{code:MB17}{https://www.github.com/una-dinosauria/human-motion-prediction}
\newurl{code:LZ18}{https://www.github.com/chaneyddtt/Convolutional-Sequence-to-Sequence-Model-for-Human-Dynamics}
\newurl{code:PG18}{https://www.github.com/facebookresearch/QuaterNet}
\newurl{code:CA19}{https://www.github.com/eddyhkchiu/pose_forecast_wacv}
\newurl{code:GM19}{https://www.github.com/cr7anand/neural_temporal_models}
\newurl{code:LiuW19}{https://www.github.com/BII-wushuang/Lie-Group-Motion-Prediction}
\newurl{code:GC19}{https://www.github.com/CHELSEA234/SkelNet_motion_prediction}
\newurl{code:KG19}{https://www.github.com/maharshi95/Pose2vec}
\newurl{code:AK19}{https://www.github.com/eth-ait/spl}
\newurl{code:ML19}{https://www.github.com/wei-mao-2019/LearnTrajDep}
\newurl{code:LC20}{https://www.github.com/limaosen0/DMGNN}
\newurl{code:ML20}{https://www.github.com/wei-mao-2019/HisRepItself}
\newurl{code:LK20}{https://www.github.com/tileb1/motion-prediction-tim}
\newurl{code:BG21}{https://www.github.com/bouracha/OoDMotion}
\newurl{code:TH06}{https://www.github.com/uoguelph-mlrg/mhmublv}
\newurl{code:HS16}{https://www.theorangeduck.com/page/deep-learning-framework-character-motion-synthesis-and-editing}
\newurl{code:ZL18}{https://www.github.com/papagina/Auto_Conditioned_RNN_motion}
\newurl{code:BK18}{https://www.github.com/ebarsoum/hpgan}
\newurl{code:YR18}{https://www.github.com/xcyan/eccv18_mtvae}
\newurl{code:HG19}{https://www.github.com/magnux/MotionGAN}
\newurl{code:WY20}{https://www.github.com/zheshiyige/Learning-Diverse-Stochastic-Human-Action-Generators-by-Learning-Smooth-Latent-Transitions}
\newurl{code:AS20}{https://www.github.com/mix-and-match/mix-and-match-tutorial}
\newurl{code:WA20}{https://www.github.com/lucaskingjade/Motion_Synthesis_Adversarial_Learning}
\newurl{code:HY20}{https://www.github.com/ubisoft/ubisoft-laforge-animation-dataset}
\newurl{code:YK20}{https://www.github.com/Khrylx/DLow}
\newurl{code:HA20}{https://www.github.com/simonalexanderson/StyleGestures}
\newurl{code:HK17}{https://www.theorangeduck.com/page/phase-functioned-neural-networks-character-control}
\newurl{code:ZS18}{https://www.github.com/ShikamaruZhang/MANN}
\newurl{code:MS18}{https://www.github.com/ianxmason/Fewshot_Learning_of_Homogeneous_Human_Locomotion_Styles}
\newurl{code:SZ19}{https://www.github.com/sebastianstarke/AI4Animation/blob/master/AI4Animation/SIGGRAPH_Asia_2019}
\newurl{code:RH17}{https://www.github.com/JooseRajamaeki/TVCG18}
\newurl{code:PB17}{https://www.github.com/xbpeng/DeepLoco}
\newurl{code:PA18}{https://www.github.com/xbpeng/DeepMimic}
\newurl{code:YT18}{https://www.github.com/VincentYu68/SymmetryCurriculumLocomotion}
\newurl{code:PK18}{https://www.github.com/akanazawa/motion_reconstruction}
\newurl{code:RH19}{https://www.github.com/JooseRajamaeki/TVCG18}
\newurl{code:AL19}{https://www.github.com/UBCMOCCA/SymmetricRL}
\newurl{code:BN19}{https://www.github.com/donamin/self-imitation-learning}
\newurl{code:PR19}{https://www.github.com/snumrl/ICC}
\newurl{code:KL20}{https://www.github.com/taesoobear/IPCDNNwalk}
\newurl{code:MT20}{https://www.github.com/deepmind/deepmind-research/tree/master/catch_carry}
\newurl{code:WG20}{https://www.github.com/facebookresearch/ScaDiver}
\newurl{code:XL20}{https://www.github.com/belinghy/SteppingStone}
\newurl{code:JV19}{https://www.github.com/jyf588/lrle}
\newurl{code:LP19}{https://www.github.com/lsw9021/MASS}
\newurl{code:ZY20}{https://www.github.com/vt-vl-lab/footskate_reducer}
\newurl{code:JL20}{https://www.github.com/DK-Jang/human_motion_manifold}
\newurl{code:SA20}{https://www.github.com/Shimingyi/MotioNet}
\newurl{code:VY18}{https://www.github.com/rubenvillegas/cvpr2018nkn}
\newurl{code:LimC19}{https://www.github.com/ljin0429/bmvc19_pmnet}
\newurl{code:KP20}{https://www.github.com/medialab-ku/retargetting-tdcn}
\newurl{code:AbL20}{https://www.github.com/DeepMotionEditing/deep-motion-editing}
\newurl{code:AW20}{https://www.github.com/DeepMotionEditing/deep-motion-editing}
\definecolor{tablerowcolor1}{RGB}{240, 245, 255}    % light blue
\definecolor{tablerowcolor2}{RGB}{255, 255, 255}    % white
\definecolor{plotcolor1}{RGB}{68, 114, 196}         % blue
\definecolor{plotcolor2}{RGB}{237, 125, 49}         % orange
\definecolor{plotcolor3}{RGB}{165, 165, 165}        % gray
\newAcronym{Bidirectional LSTM}{BiLSTM}
\newAcronym{Convolutional Neural Network}{CNN}
\newAcronym{Conditional RBM}{CRBM}
\newAcronym{Discrete Cosine Transform}{DCT}
\newAcronym{Deep Learning}{DL}
\newAcronym{Deep Neural Network}{DNN}
\newAcronymPlr{Degree of Freedom}{Degrees of Freedom}{DOF}
\newAcronym{Deep Reinforcement Learning}{DRL}
\newAcronym{Encoder-Recurrent-Decoder}{ERD}
\newAcronym{Factored CRBM}{FCRBM}
\newAcronymPlr{Forward Kinematics}{Forward Kinematics}{FK}
\newAcronym{Generative Adversarial Network}{GAN}
\newAcronym{Graph Convolutional Network}{GCN}
\newAcronym{Gaussian Mixture Model}{GMM}
\newAcronym{Gated Recurrent Unit}{GRU}
\newAcronym{Hierarchical FCRBM}{HFCRBM}
\newAcronymPlr{Inverse Kinematics}{Inverse Kinematics}{IK}
\newAcronymPlr{Long Short-Term Memory}{Long Short-Term Memory}{LSTM}
\newAcronym{Markov Chain Monte Carlo}{MCMC}
\newAcronym{Machine Learning}{ML}
\newAcronym{Natural Language Processing}{NLP}
\newAcronym{Proportional-Derivative}{PD}
\newAcronym{Phase-Functioned Neural Network}{PFNN}
\newAcronym{Restricted Boltzmann Machine}{RBM}
\newAcronym{Reinforcement Learning}{RL}
\newAcronym{Recurrent Neural Network}{RNN}
\newAcronymPlr{Sequence to Sequence}{Sequence to Sequence}{Seq2Seq}
\newAcronym{Variational Autoencoder}{VAE}
\title{A Survey on Deep Learning for Skeleton-Based Human Animation}
\author[%
    L. Mourot, L. Hoyet, F. Le Clerc, F. Schnitzler~\& P. Hellier%
]{\parbox{\textwidth}{\centering%
    L. Mourot$^{1,2}$\orcid{0000-0001-8441-892X}, %
    L. Hoyet$^{1}$\orcid{0000-0002-7373-6049}, %
    F. Le Clerc$^{2}$\orcid{0000-0003-0519-8581}, %
    F. Schnitzler$^{2}$ \orcid{0000-0003-1304-2157} and %
    P. Hellier$^{2}$ \orcid{0000-0003-3603-2381}%
}\\{\parbox{\textwidth}{\centering%
    $^1$Inria, Univ Rennes, CNRS, IRISA\\
    $^2$InterDigital, Inc%
}}}

\begin{document}
    \maketitle
    
    \begin{abstract}
    Human character animation is often critical in entertainment content production, including video games, virtual reality or fiction films. To this end, deep neural networks drive most recent advances through deep learning and deep reinforcement learning. In this article, we propose a comprehensive survey on the state-of-the-art approaches based on either deep learning or deep reinforcement learning in skeleton-based human character animation. First, we introduce motion data representations, most common human motion datasets and how basic deep models can be enhanced to foster learning of spatial and temporal patterns in motion data. Second, we cover state-of-the-art approaches divided into three large families of applications in human animation pipelines: motion synthesis, character control and motion editing. Finally, we discuss the limitations of the current state-of-the-art methods based on deep learning and/or deep reinforcement learning in skeletal human character animation and possible directions of future research to alleviate current limitations and meet animators' needs.


\end{abstract}
    \section{Introduction}
    % General introduction
    Humans and their representations are ubiquitous in culture. In the past decades digital technologies brought more and more tools to assist designers, animators and artists, with the goal of increasing their creative capabilities and the realism of their artworks while reducing production costs. For instance, digital doubles have brought to life non-human fictional creatures like the Na'vi in \emph{Avatar} or even visual identity of long time deceased actors like Grand Moff Tarkin in \emph{Rogue One} in 2016 portrayed by the British actor Peter Cushing, even though he passed away in 1994. Within this context, computer-assisted human character animation plays a central role, such as for the synthesis of the motion of digital doubles. Another example is the video game production, where animation and control of the characters directly condition the success of a game.
    
    % Challenges
    However, animation of human characters is challenging: the way humans move is very diverse and is influenced by many factors including the mood, the intentions, the activity or even individual characteristics. In addition, Newton's second law of motion inherently makes human motion a dynamic process, while our understanding of biomechanics is far from being comprehensive. In practice, compromises therefore need to be made to balance between production costs, the realism, the amount of manual work, and the level of expertise of the animator. The research area of character animation has been active for decades to mitigate these compromises and make animation more accessible, starting from pioneering works such as editing and deforming motion examples~\cite{ref:WP95}, retargeting motions to new characters~\cite{ref:Gl98}, controlling characters using motion graphs~\cite{ref:KG02, ref:MC12}, etc.

    % DNNs introduction
    Over the past few years, \acrFullPlr{DNN} have emerged as a powerful means to enhance the performance and capabilities of character animation, as evidenced by the growing number of publications on the topic leveraging \acrFull{DL} and \acrFull{DRL} (see Figure~\ref{fig:papers}). These two techniques have shown an unprecedented ability to address complex tasks in a wide variety of domains not restricted to animation, such as computer vision, \acrFull{NLP} and many more. Humans are outperformed by artificial intelligence algorithms in a growing number of tasks such as image classification or playing Go. This is notably due to the fact that \acrShortPlr{DNN} are powerful function approximators, able to learn sophisticated patterns in complex real-world phenomena from data. Moreover, once the training is complete, training data is discarded, leaving compact models that are able to meet performance requirements of real-time applications or to scale to embedded systems. In animation, \acrShort{DL} \& \acrShort{DRL} based approaches attempt to handle the human motion complexity and provide promising perspectives for cheaper and faster animation techniques with more and more fidelity and creative capabilities.
    
    % Article introduction
    In this article, we present an overview of the recent growing trend of \acrShort{DL} \& \acrShort{DRL} in skeletal character animation, focused on humanoid characters. \emph{Skeletal} here means using a representation derived from a skeleton, as commonly used in the movie and game industries in combination with \DDD skinned meshes (see Section~\ref{subsection:PoseRepr}). On the one hand, \acrShort{DL} is particularly effective at building compact models from such human motion data, \eg for motion synthesis and editing. On the other hand, \acrShort{DRL}, which is concerned with how agents ought to act in a simulated environment in order to maximise some cumulative reward, is well-suited for character control where characters are agents, and their actuation model parameters are actions. The goal of this review is therefore to provide researchers involved in character animation and related fields with an overview of existing \acrShort{DL} \& \acrShort{DRL} based methods in that domain, best practices to represent and process motion data with \acrShortPlr{DNN} and the most successful approaches for different well-studied families of problems.

    % Related works
    \pgfkeys{/pgf/number format/.cd,1000 sep={}}
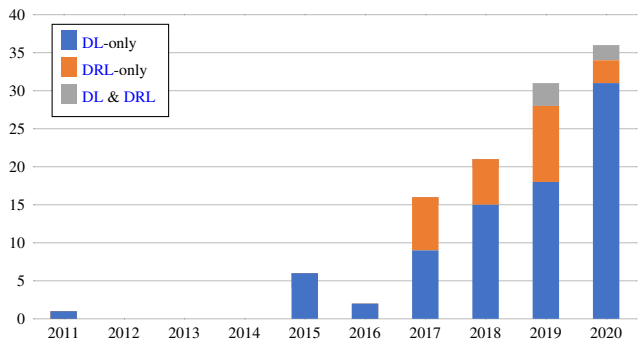
\begin{figure}[t]
    \centering
    \begin{tikzpicture}
        \begin{axis}[
            ybar stacked,
            %
            % X axis, ticks & labels
            x=0.095\linewidth,
            xmin=2010.5, xmax=2020.5,
            xtick={2011, 2012, 2013, 2014, 2015, 2016, 2017, 2018, 2019, 2020},
            xticklabel style={font=\tiny},
            %
            % Y axis, ticks & labels
            y=0.012\linewidth,
            ymin=0, ymax=40,
            ytick={0, 5, 10, 15, 20, 25, 30, 35, 40},
            yticklabel style={font=\tiny},
            %
            % Global axis & grids
            axis line style={draw=none},
            major tick length=0,
            ymajorgrids,
            xmajorgrids=false,
            %
            % Legend
            legend style={font=\tiny},
            legend pos=north west,
            legend cell align={left},
        ]
        
            \addplot[draw opacity=0.0, fill=plotcolor1] coordinates {
                (2011, 1)
                % (2012, 0)
                % (2013, 0)
                % (2014, 0)
                (2015, 6)
                (2016, 2)
                (2017, 9)
                (2018, 15)
                (2019, 18)
                (2020, 31)
            };
            \addlegendentry{\acrShort{DL}-only}
            
            \addplot[draw opacity=0.0, fill=plotcolor2] coordinates {
                (2011, -1)
                % (2012, 0)
                % (2013, 0)
                % (2014, 0)
                (2015, -1)
                (2016, -1)
                (2017, 7)
                (2018, 6)
                (2019, 10)
                (2020, 3)
            };
            \addlegendentry{\acrShort{DRL}-only}
            
            \addplot[draw opacity=0.0, fill=plotcolor3] coordinates {
                (2011, -1)
                % (2012, 0)
                % (2013, 0)
                % (2014, 0)
                (2015, -1)
                (2016, -1)
                (2017, -1)
                (2018, -1)
                (2019, 3)
                (2020, 2)
            };
            \addlegendentry{\acrShort{DL} \& \acrShort{DRL}}
        \end{axis}
    \end{tikzpicture}
    \caption{
        Histogram of the peer-reviewed publications over the past decade in human skeletal character animation using \acrFull{DL} and/or \acrFull{DRL} addressed in this survey.
    }
    \label{fig:papers}
\end{figure}
    In the last decade, various state-of-the-art reports have studied character animation topics: Pejsa and Pandzic~\cite{ref:PP10} covered motion planning, motion graphs and parametric models for motion synthesis in interactive applications from examples; Geijtenbeek and Pronost~\cite{ref:GP12} reviewed the literature on physical simulation for interactive character animation; then Karg~\etal~\cite{ref:KS13} studied the generation and recognition of motions based on affective expressions; finally, Wang~\etal~\cite{ref:WC14} presented an analysis of state-of-the-art techniques in \DDD human motion synthesis while focusing on motion capture data-driven methods. However, none of these reviews covered \acrShort{DL} or \acrShort{DRL} based methods because most recent advances appeared since 2015, as shown in Figure~\ref{fig:papers}. More recently, Alemi and Pasquier~\cite{ref:AP19} reviewed the topic of data-driven movement generation with \acrFull{ML}. While their paper includes the review of some \acrShort{DL}-based methods, numerous other advances have emerged in very recent years. In addition, some of the \acrShort{ML} approaches presented by Alemi and Pasquier~\cite{ref:AP19}, such as Hidden Markov Models or Principal Component Analysis, are clearly outperformed by \acrShort{DL} as of today. In contrast, our work addresses state-of-the-art of \acrShort{DL} and \acrShort{DRL} in skeletal human character animation.

    % Paper organization: section 2
    This paper is organised as follows: in Section~\ref{section:MotionRepr&Modelling}, we begin with an overview of low-level concerns encountered when processing human motion data with \acrShortPlr{DNN}, presenting pose representations (Section~\ref{subsection:PoseRepr}) and human motion datasets (Section~\ref{subsection:Datasets}) frequently used in the literature, as well as how to efficiently and successfully learn spatial (Section~\ref{subsection:SpatialFeatures}) and temporal (Section~\ref{subsection:TemporalFeatures}) features. 
    
    % Paper organization: section 3
    Next, Section~\ref{section:MotionSynthesis} covers motion synthesis, that we define as the process of creating perceptually plausible motion sequences with a desired style or expressed emotion for instance. Motion synthesis models are thus capable of generating different motions depending on inputs, including low-level parameters (\eg, latent variables), high-level parameters (\eg trajectory or specific action to perform) or historical parameters (\eg past motions to be extrapolated). In this review, we divide motion synthesis approaches into three categories: predictive in the short-term (Section~\ref{subsection:ShortTermPrediction}), predictive in the long-term (Section~\ref{subsection:LongTermPrediction}) and generative (Section~\ref{subsection:DeepGenerativeModeling}). The first focuses on deterministically synthesising motion segments from few past frames while the second tries to generate longer plausible motion continuations. The third category covers motion synthesis from other types of parameters, such as the trajectory to follow, and relies on generative models.
    
    % Paper organization: section 4
    Section~\ref{section:CharCtrl} deals with the task of controlling the motion of characters so that they react naturally to user inputs while accounting for environment and biomechanical constraints. We divide these approaches into kinematic (Section~\ref{subsection:KinCharCtrl}), physical (Section~\ref{subsection:PhysCharCtrl}) and biomechanical (Section~\ref{subsection:BioCharCtrl}) control. Kinematic approaches directly produce motions as joint positions or angles. In contrast, both physical and biomechanical approaches strive to obey the laws of physics, while differing in the actuation model: physical models are actuated by forces and torques, while biomechanical models (a.k.a. musculoskeletal models) are driven by muscle activations.
    
    % Paper organization: section 5
    Finally, Section~\ref{section:MotionEditing} gathers motion editing approaches at large, \ie methods aiming to process or transform some aspects of existing motion data. Motion cleaning (Section~\ref{subsection:Cleaning}) improves motion data, \eg by removing noise or filling in missing information such as marker or joint positions. Note that we distinguish here markers and joints completion from motion prediction and in-betweening that are addressed in Sections~\ref{subsection:ShortTermPrediction} and~\ref{subsection:LongTermPrediction} respectively, although all three might be formulated as completion from partial observations. Then, retargeting (Section~\ref{subsection:Retargeting}) strives to transfer the motion from a source character to a target character, while motion style transfer (Section~\ref{subsection:StyleTransfer}) edits the style of a motion segment while preserving the action performed and the character.
    \section{Human Motion Representation, Data and Modelling}
\label{section:MotionRepr&Modelling}
    In both \acrShort{DL} and \acrShort{DRL}, choices of input and output spaces for \acrShortPlr{DNN} are often impactful on the effectiveness of the learning and on what specific aspects of the data will be retained. When dealing with human motion, the pose representation mainly determines these input and output spaces. Moreover in \acrShortPlr{DNN}, the computational workflow can be structured around spatial and temporal aspects of motion. In this section we explore the different human pose representations commonly encountered in deep animation and their key strengths and weaknesses (Section~\ref{subsection:PoseRepr}), commonly used datasets (Section~\ref{subsection:Datasets}), as well as \acrShort{DNN} architectures structured with respect to spatial (Section~\ref{subsection:SpatialFeatures}) and temporal (Section~\ref{subsection:TemporalFeatures}) domains.

\subsection{Pose Representation}
\label{subsection:PoseRepr}
    Traditional animation approaches typically use \DDD rigged skeletons with skinned meshes, which provides a good trade-off between quality and complexity. Rigging offers convenient ways to manipulate \DDD models as strings do for a puppet, while skinning is the process of binding actual \DDD meshes to animated characters. In that framework, human motions are usually represented as sequences of poses separated by constant time intervals whose rate generally ranges from 30 to 250 hertz. At each time step, the state of the human body is then represented as a set of links (\ie bones) connected by joints. This skeletal representation is a good compromise for the complexity and the diversity of human movements that can be represented. When bone lengths are kept constant over time, the \acrFullPlr{DOF} are the orientations of the bones, commonly expressed relative to their parent. In the following, we call such a representation an \emph{angular pose representation}, as opposed to a \emph{positional pose representation} where the skeleton \acrShortPlr{DOF} are the coordinates of the joint positions, which does not explicitly constrain bone lengths to remain constant over time.
    
    \subsubsection{Positional Pose Representations}
        In a positional pose representation, each joint is directly represented by its position, generally expressed at each time step in the body's local coordinate system~\cite{ref:HS15, ref:HS16, ref:WH19, ref:HA20, ref:DH19, ref:HH17, ref:ZL18, ref:HoH17, ref:SC19, ref:TC17, ref:KA20}, which allows the decomposition of the whole motion into the local movements of limbs with respect to the body itself and the global movement of the body with respect to its environment. Although different coordinate systems could be formulated to embed the set of joint positions, positional pose representations almost always rely on the Cartesian coordinate system. It has neither discontinuities nor singularities and constitutes a convenient space for interpolation, visualisation and optimisation. Moreover, within the framework described here, positional pose representation does not present some of the limitations inherent to angular representations presented in the following section. However, it suffers from some limitations related to the structure of human motions. For instance, joint positions do not encode the information of bone orientations around themselves which is often needed for concrete applications in animation, in order to display more natural mesh deformations. Positional representations also do not explicitly constrain bone lengths to remain constant over time, therefore requiring reliance on additional constraints to ensure that the skeleton does not break apart. For these reasons, communities closer to animation, such as computer graphics, rarely use purely positional pose representations, while on the opposite, communities more commonly involved in deep learning, such as computer vision, are more prone to employing these representations.
    
    \subsubsection{Angular Pose Representations}
    \label{subsection:AngPoseRepr}
        Angular representations have been widely used in animation mainly because their hierarchical nature allows straightforward orientation of any joint together with all of its descendants, while keeping bone lengths constant. Indeed, the position of each joint is described with respect to its parent as a \DDD rigid transformation, often decomposed into a variable rotation and a fixed translation, corresponding to the joint orientation and the bone dimensions respectively. Main differences among angular pose representations are determined by the parameterisation of the rotations, however there are also representations working at the level of rigid transformation parameterisations.

        Formally, the set of all rotations of $\R{3}$ equipped with the composition is the \DDD rotation group often denoted $\SO{3}$, standing for special orthogonal group of dimension 3. $\SO{3}$ can be identified with the group of orthogonal $3 \times 3$ matrices with determinant~1 under the matrix multiplication. %
        Similarly, the \DDD special Euclidean group whose elements are proper \DDD rigid transformations (\ie excluding reflections) is $\SE{3} = \SO{3} \times \R{3}$. Both $\SO{3}$ and $\SE{3}$ are Lie groups, \ie differentiable spaces that locally resemble Euclidean space. Furthermore, a Lie algebra is associated to every Lie group, called $\so{3}$ and $\se{3}$ for $\SO{3}$ and $\SE{3}$, respectively. Lie algebras are vector spaces tangent to their Lie group at the identity element completely capturing its local structure, making them compelling as representation spaces.
        
        \paragraph*{Euler Angles.}
        The most intuitive parameterisation of $\SO{3}$ is probably \emph{Euler angles}, that represents a \DDD orientation as three successive rotations around different axes, \eg yaw, pitch and roll. However, it suffers from the well-known gimbal lock when two of the three rotation axes align, causing a \acrShort{DOF} to be lost. Gimbal lock can be avoided only if at least one rotation axis is limited to a range smaller than $180^\circ$, which is not always possible in practice. As a result, Euler angles are unsuitable for \acrFull{IK}, dynamics and spacetime optimisation~\cite{ref:G98}. Moreover, they do not work well for interpolations since the space of orientations is highly nonlinear~\cite{ref:G98}. Finally, multiple conventions exist for the axes considered, including their order, which requires to define them formally in each application to avoid any ambiguity. For these reasons, this representation is inappropriate for a lot of applications.

        \paragraph*{Lie Algebras.}
        A popular angular pose representation in skeletal character animation consists in representing each joint rotation as an element of $\so{3}$, where the direction and magnitude of the vector correspond to the axis and angle of the rotation~\cite{ref:G98}, respectively. Since such a vector is an element of $\so{3}$, the parameterisation is defined by the exponential map from $\so{3}$ to $\SO{3}$ which can be efficiently computed with the Rodrigues' formula~\cite{ref:R40}. This pose representation is often called \emph{exponential map}, and is sometimes confused with the so-called \emph{axis-angle} representation that is equivalent but separates the vector into a unit vector and a scalar describing the axis and the magnitude of the rotation.
        
        Since Lie algebras are locally linearised versions of their Lie group, $\so{3}$ is a compelling space to work with elements of $\SO{3}$. However, as all parameterisations of $\SO{3}$ in $\R{3}$, the exponential map representation has singularities~\cite{ref:G98} leading to losing a \acrShort{DOF} in some parts of the representation space, even though these are located on the spheres of radius $2k\pi$ for $k \in \mathbb{N}^+$, since a rotation of $2\pi$ about any axis is equivalent to no rotation. Therefore, this representation is often well-suited in animation since control and simulation deal with small time steps and thus with small rotations that stay inside the sphere of radius $2\pi$, far from the singularities. It has been employed in early works in deep animation~\cite{ref:TH06, ref:TH09}, and broadly exploited for motion synthesis~\cite{ref:CM11, ref:AL15} and prediction~\cite{ref:FL15, ref:JZ16, ref:BB17, ref:MB17, ref:LZ18, ref:TM18, ref:GW18a, ref:GM19, ref:LiW19, ref:GC19, ref:KG19, ref:WA19, ref:CS20, ref:ML20, ref:LC20, ref:CS21, ref:LC21}, as well as in other topics~\cite{ref:HK17, ref:AP17, ref:MS18, ref:JL20}. However, as quality needs increase, long-term correlations are more and more important and inevitably imply larger rotations, getting close to the singularities of the parameterisation.
        
        Similarly to the exponential map representation which uses $\so{3}$ to represent joint orientations w.r.t. their parents, Liu~\etal~\cite{ref:LiuW19} proposed a pose representation using $\se{3}$ to represent rigid transformations of each joint w.r.t. its parent. The main motivation for choosing such a representation is to explicitly encode both geometric constraints (\ie bone lengths) and actual \acrShortPlr{DOF} (\ie joint orientations) together. Nevertheless, it still has the same singularities as the exponential map representation. As we will see in the following paragraphs, other parameterisations in higher-dimensional spaces than $\R{3}$, \ie over-parameterised representations, are able to prevent such singularities.

        \paragraph*{Rotation Matrices.}
        In computer graphics, rotation matrices are widely used to represent \DDD rotations. The corresponding parameterisation is the identity since elements of $\SO{3}$ are $3 \times 3$ matrices. Rotation matrices have no singularity and can be integrated together with joint translations into a $4 \times 4$ homogeneous matrix, which is elegant and effective when involved in computations like composition or inverse. However, such a representation is particularly difficult to work with when its parameters must be estimated since the representation is over-parameterised. Indeed, not all $3 \times 3$ matrices belong to $\SO{3}$. By definition, a matrix $R \in \R{3 \times 3}$ must satisfy $R^{\top}R=I$ and $det(R) = 1$ to be a valid \DDD rotation. For instance, predicting the orientation of a joint in matrix representation would require to solve a constrained optimisation problem to ensure the validity of the rotation, which can be tedious.

        \paragraph*{Unit Quaternions.}
        A more compact representation than rotation matrices are unit quaternions. Lying in $\R{4}$, they are free of singularities, suitable for interpolation~\cite{ref:G98}, numerically stable and computationally efficient~\cite{ref:PG18}. Like $\so{3}$, the space of unit quaternions has the same local geometry and topology as $\SO{3}$~\cite{ref:G98}. However, unit quaternions are also over-parameterised, but have only four parameters (in comparison to nine parameters for rotation matrices). Thus, they must be constrained to remain on the unit 4-sphere. This angular representation has been popularised in \acrShort{DL}-based animation with \emph{QuaterNet}, a quaternion-based framework for human motion prediction proposed by Pavllo~\etal~\cite{ref:PG18, ref:PF20} (see Section~\ref{subsection:ShortTermPrediction}). In that framework, Pavllo~\etal introduced a penalty term in the loss function for all quaternions predicted by the network that minimises their divergence from the unit length. It encourages the network to predict valid rotations and leads to better training stability. Moreover, the predicted quaternions are also normalised after computing the penalty to enforce their validity. According to the authors, the distribution of predicted quaternion norms converges to a Gaussian with mean 1 during the training, suggesting that the model actually learns to represent valid rotations. Since Pavllo~\etal~\cite{ref:PG18} showed promising results using quaternions, their use is gaining popularity~\cite{ref:KP20, ref:AW20, ref:AS20, ref:AbL20, ref:PF20, ref:LimC19, ref:VY18, ref:HY20, ref:GW20}.

        \paragraph*{Gram-Schmidt-like.}
        Zhou~\etal~\cite{ref:ZB19} recently pointed out that all representations in $\R{n}$ with $n \leq 4$ have discontinuities which can be unfavorable for \acrShortPlr{DNN} training. They therefore introduced a continuous representation of $n$-dimensional rotations $\SO{n}$ with $n^2-n$ dimensions. The mapping from $\SO{n}$ to the representation space simply drops the last column vector of the input $n \times n$ matrix. The inverse mapping back to $\SO{n}$, called Gram-Schmidt-like process, is a Gram-Schmidt process over the $n-1$ column vectors followed by the computation of the last column vector by a generalisation to $n$ dimensions of the cross product. In the case of $\SO{3}$, this \emph{Gram-Schmidt-like} representation gives a \nD{6} representation. Zhou~\etal~\cite{ref:ZB19} also provided a method to further reduce the dimensionality from \nD{6} to \nD{5} while still keeping a continuous representation using a stereographic projection combined with normalisation. However, they empirically found that nonlinearities introduced by the projection can make the learning process more difficult. To the best of our knowledge, very few works have investigated this promising representation of \DDD rotations.

        \begin{figure}[h]
    \centering
    \includegraphics[width=0.73\linewidth]{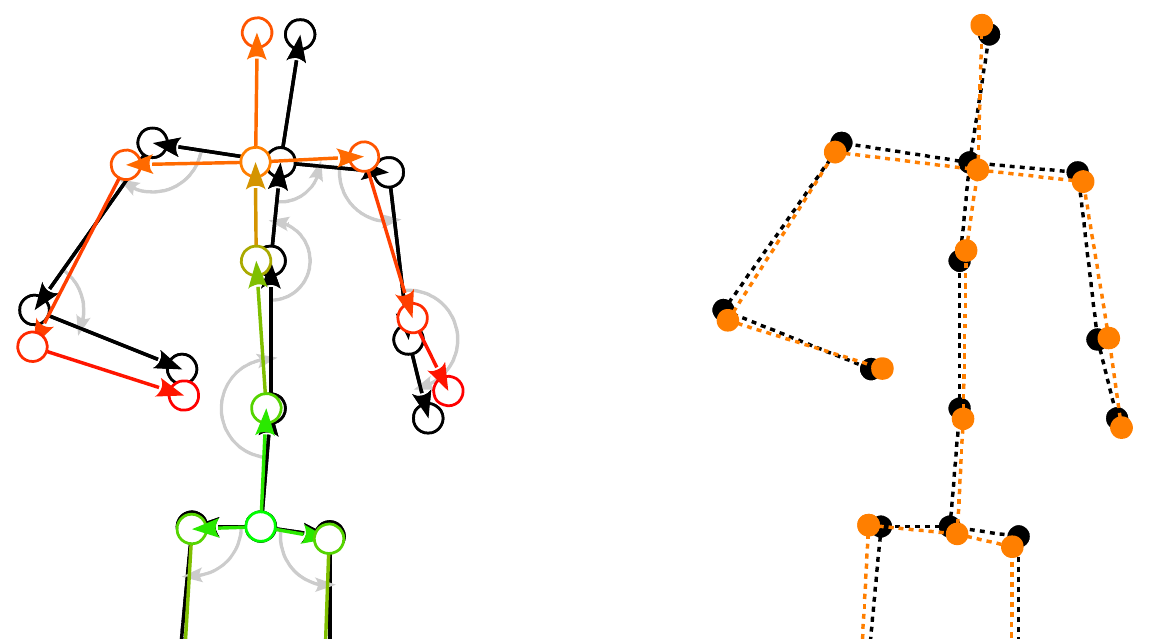}
    \caption{
        Illustration of the error accumulation problem with angular representations: even small angular errors along the kinematic chain can lead to large accumulated joint positioning errors (left-hand stick figure, see color gradient). This is problematic in optimization-based methods, \eg \acrShort{DL}/\acrShort{DRL}, when penalizing joint orientation deviations. This is not the case with positional representations (right-hand stick figure) where joint positions are directly optimized.
    }
    \label{fig:error_accumulation}
\end{figure}
        \paragraph*{Hierarchical Representation Limitations.}
        % Intrinsic problem of angular representation
        In skeletal character animation, a hierarchical modelling approach is used in conjunction with angular pose representations, \ie the representation of the joint orientations relative to their parents. In that context, positional errors over proximal joints (\eg the shoulder) are propagated and accumulated down the kinematic chains. This is problematic in optimisation-based methods such as \acrShort{DL} or \acrShort{DRL} since equally distributed joint orientation errors will result in growing joint position errors along the kinematic chains as depicted in Figure~\ref{fig:error_accumulation}, making it difficult to accurately handle end effectors. This is especially true in motions sequences involving fast or ample movements, \eg running.
        
        % Pavllo's forward kinematics approach
        To solve this problem, Pavllo~\etal~\cite{ref:PG18, ref:PF20} performed \acrFull{FK} to convert quaternion-based poses predicted by their \acrShort{DNN} into \DDD joint positions, and then penalised absolute position errors instead of angular errors. Since \acrShort{FK} is a differentiable operation with respect to joint orientations, they can train their network end-to-end using a positional loss.
        
        % Ghorbani's joints weighting approach
        While recent works have followed the same approach~\cite{ref:AS20, ref:AbL20}, Ghorbani~\etal~\cite{ref:GW20} recently pointed out that applying \acrShort{FK} is computationally expensive especially for motion sequences that are long or involve numerous joints. To this end, they proposed to hierarchically weight joint angle errors based on their impact on the positions. They set the weight of a joint as its maximum path length down to all of the connected end effectors in an average skeleton. While ablative studies showed improved performances when joint weighting was enabled, the results were not compared to Pavllo~\etal's \acrShort{FK}-based approach~\cite{ref:PG18, ref:PF20}, neither were the choice of the weights assessed.

    \subsubsection{Hybrid Representations}
        As mentioned above, both angular and positional approaches for representing human poses have advantages and drawbacks that sometimes depend on the application or viewpoint, dividing researcher communities. For this reason, several works have proposed hybrid representations with the goal of mitigating drawbacks while keeping benefits of both types of representations.

\begin{figure}[!ht]
    \centering
    \includegraphics[width=\linewidth]{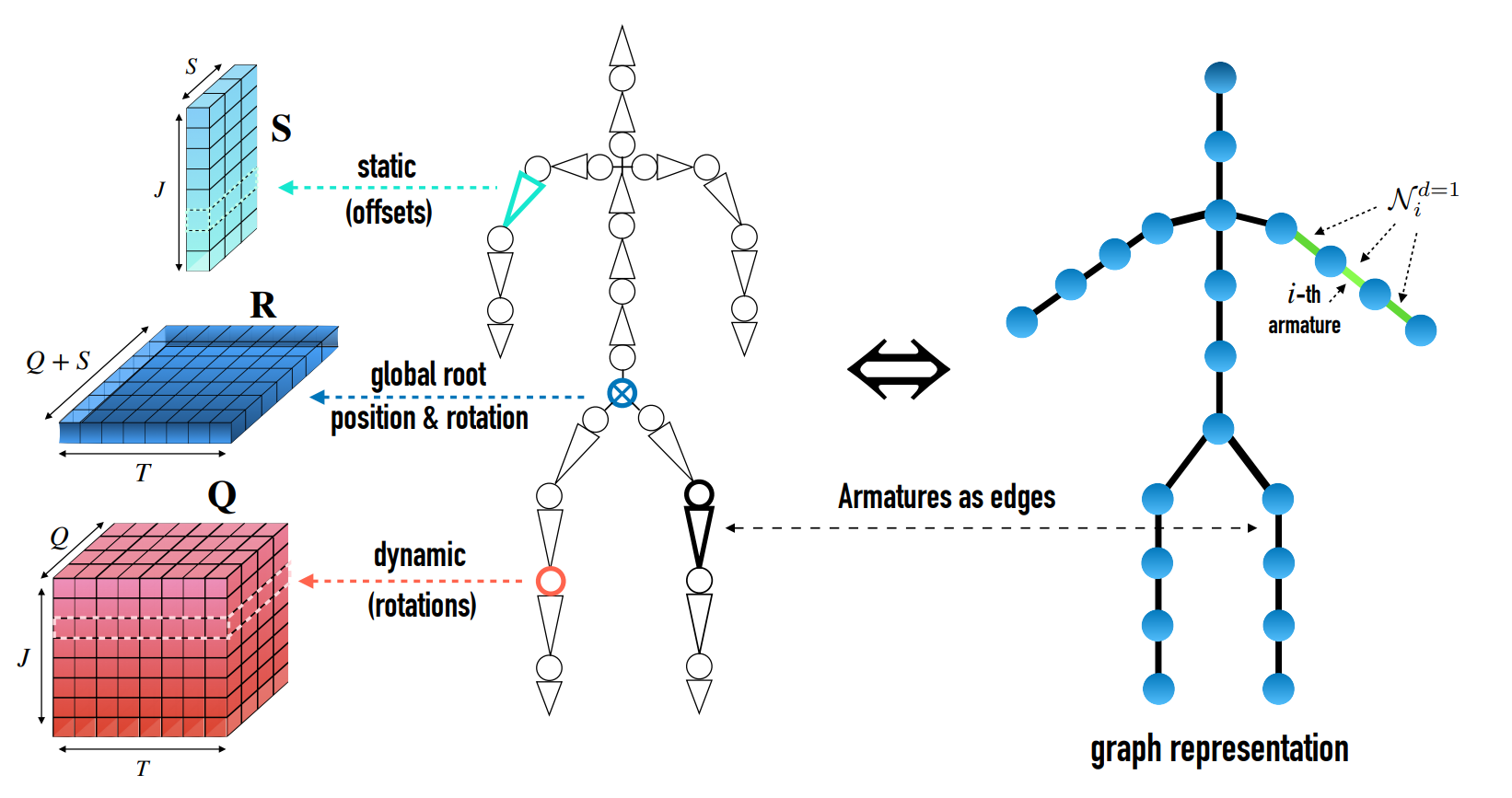}
    \caption{
        Representation of skeletal motion data as a graph in~\cite{ref:AbL20}. The nodes of the graph correspond to joints and the edges to armatures. Each of the $J$ armatures holds a time-varying tensor $Q$ modelling the temporal sequence of rotations at its corresponding joint, and a time-independent vector $S$ modelling the bone offset to the parent joint. The global motion of the root joint $R$ is processed separately. Figure from~\cite{ref:AbL20}.
    }
    \vspace{0.5cm}
    \label{fig:AbL20}
\end{figure}

        % Aberman's hybrid representation
        Aberman~\etal~\cite{ref:AbL20} presented a novel data-driven approach for retargeting motions between homeomorphic skeletons (see Section~\ref{subsection:Retargeting}) along with an interesting and elegant representation of human motion illustrated in Figure~\ref{fig:AbL20}. In this work, both angular and positional information are combined: a static component~$S$ consisting of a set of \DDD positional offsets describes the skeleton in some arbitrary pose (similar to a T-pose but specific to a pose sequence), while a dynamic component $Q$ specifies the sequence of orientations of each joint along time (with respect to $S$) represented using unit quaternions. The separation between static and dynamic partial representations enables the authors to design an architecture such that each component is processed in a separate branch. In continuation of Pavllo~\etal's work~\cite{ref:PG18, ref:PF20}, Aberman~\etal~\cite{ref:AbL20} penalised errors in the positional space after performing \acrShort{FK}, while also penalising errors in the quaternion space. Following this work, Shi~\etal~\cite{ref:SA20} addressed the reconstruction of \DDD kinematic skeletons from \DD keypoints estimated from monocular video while dividing the motion representation into static bone lengths and dynamic joint orientations. To this end, a \acrShort{DNN} called \emph{MotioNet} learns to map \DD keypoints to a symmetric static skeleton, represented by its bone lengths and a dynamic sequence of joint rotations (quaternions) which are then combined through \acrShort{FK} to get a full kinematic skeleton.
        
        % Redundant representations
        Finally, the success of multiple recent methods in skeletal character animation mixing different pose representations suggests that \acrShortPlr{DNN} benefit from redundant pose information. In addition to joint orientations, researchers fed their models with joint positions~\cite{ref:LL18}, joint positions and velocities~\cite{ref:HK17, ref:MS18, ref:ZS18, ref:LingZ20, ref:SZ20, ref:SZ19} or even joint positions and linear and angular velocities~\cite{ref:HK20}.

\subsection{Human Motion Datasets}
\label{subsection:Datasets}
\newcommand{\sizeEntry}[3]{%
    #1 frames @ \makebox[1.1cm][l]{#2 Hz} $\rightarrow$ \makebox[0.6cm][r]{#3 h}%
}
\begin{table*}
    \setlength{\tabcolsep}{12.5pt}
    \scriptsize
    \centering
    \caption{
        Summary of the main datasets presented in Section~\ref{subsection:Datasets}.
    }
    \begin{NiceTabular}{| c | c | c | *{3}{c} | c |}
        \CodeBefore
            \rowcolors{3}{tablerowcolor2}{tablerowcolor1} % Set alternating row colors
        \Body
        \hline
        & & & & & & \\
        
        % Headers
        \header{Dataset} &
        \header{URL} &
        \header{Size} &
        \Block{1-3}{\header{Data}} &&&
        \header{Availability} \\
        
        % Sub-headers
        & % no sub-header (Dataset)
        & % no sub-header (URL)
        & % no sub-header (Size)
        \subheader{Joints} &
        \subheader{Representation} &
        \subheader{Miscalleneous} &
        \\ % no sub-header (Availability)
        
        \hline
        CMU \cite{ref:CM03}             & \useurl{data:CM03} & \sizeEntry{$3.9*10^6$}{$120$}{$9.1$}        & 29 & angular      &           & Public        \\
        HDM05 \cite{ref:MR07}           & \useurl{data:MR07} & \sizeEntry{$3.6*10^5$}{$120$}{$0.8$}        & 31 & angular      &           & Public        \\
        Human3.6M \cite{ref:IP14}       & \useurl{data:IP14} & \sizeEntry{$3.6*10^6$}{$50$}{$20.0$}        & 32 & angular      & RGB+D     & On request    \\
        Holden \etal \cite{ref:HS16}    & \useurl{data:HS16} & \sizeEntry{$6.0*10^6$}{$120$}{$13.9$}       & 21 & positional   &           & Public        \\
        NTU RGB+D \cite{ref:SL16}       & \useurl{data:SL16} & \sizeEntry{$4.0*10^6$}{$30$}{$37.0$}        & 25 & positional   & RGB+D+IR  & On request    \\
        NTU RGB+D 120 \cite{ref:LS20}   & \useurl{data:LS20} & \sizeEntry{$8.0*10^6$}{$30$}{$74.1$}        & 25 & positional   & RGB+D+IR  & On request    \\
        3DPW \cite{ref:VH18}            & \useurl{data:VH18} & \sizeEntry{$5.1*10^4$}{$30$}{$0.5$}         & 23 & angular      & RGB       & Public        \\
        AMASS \cite{ref:MG19}           & \useurl{data:MG19} & \sizeEntry{$1.8*10^7$}{$[60, 250]$}{$41.5$} & 52 & angular      & Body mesh & Public        \\
        Mixamo \cite{ref:MI21}          & \useurl{data:MI21} & \sizeEntry{$2.7*10^5$}{$30$}{$2.5$}         & 52 & angular      & Body mesh & Public        \\
        \hline
    \end{NiceTabular}
    \label{tab:datasets}
\end{table*}

    % introduction
    Another crucial aspect of data-driven approaches is the choice of dataset, from which a \acrShort{DL}-based model will learn a deep representation of human motion. In particular, large amounts of high-quality motion data are necessary to constitute so-called benchmark datasets and to provide robust assessment procedures. In this section we introduce a selection of human motion datasets that are the most relevant for this survey. Although many datasets have been proposed, only a small number of them have been repeatedly exploited and even fewer can be considered as standard benchmarks (see Figure~\ref{fig:datasets}). Table~\ref{tab:datasets} provides relevant information about these datasets most commonly used in the works presented in this survey.

    % CMU & Human3.6
    The two most widely used databases in skeleton-based deep human animation are CMU~\cite{ref:CM03} and Human3.6M~\cite{ref:IP14}. Both are standard large-scale human motion datasets for learning and evaluation. Despite the fact that the CMU dataset was released about a decade earlier than Human3.6M, they have a comparable size (see Table~\ref{tab:datasets}). The main advantage of Human3.6M over CMU is the presence of RGB+D videos synchronized with human pose sequences, making it sometimes more suitable for tasks closer to computer vision such as motion prediction, even though CMU is also often leveraged. To obtain convincing results, several researchers provide evaluations of their work over both CMU and Human3.6M data~\cite{ref:BB17, ref:LZ18, ref:LiW19, ref:GC19, ref:KG19, ref:ML19, ref:CS20, ref:LC20, ref:ZP20, ref:CH20, ref:LK20, ref:LC21, ref:CS21, ref:BG21, ref:AS20}.
    
    \newcommand{\xlabel}[1]{%
    \rotatebox{90}{#1}%
}

\begin{figure}
    \centering
    \begin{tikzpicture}
        \begin{axis}[
            ybar interval,
            %
            % X axis, ticks & labels
            x=0.0325\linewidth,
            xmin=1, xmax=30,
            xticklabels={
            	\xlabel{Human3.6M \cite{ref:IP14}},
            	\xlabel{CMU \cite{ref:CM03}},
            	\xlabel{Holden \etal \cite{ref:HS16}},
            	\xlabel{NTU RGB+D \cite{ref:SL16}},
            	\xlabel{3DPW \cite{ref:VH18}},
            	\xlabel{HDM05 \cite{ref:MR07}},
            	\xlabel{Mixamo \cite{ref:MI21}},
            	\xlabel{Xia \etal \cite{ref:XW15}},
            	\xlabel{SFU \cite{ref:SF19}},
            	\xlabel{AMASS \cite{ref:MG19}},
            	\xlabel{Alemi \etal \cite{ref:AL15}},
            	\xlabel{Holden \etal \cite{ref:HK17}},
            	\xlabel{Hsu \etal \cite{ref:HP05}},
            	\xlabel{Emilya \cite{ref:FP14}},
            	\xlabel{MHAD \cite{ref:OC13}},
            	\xlabel{Habibie \etal \cite{ref:HH17}},
            	\xlabel{DeepCap \cite{ref:HX20}},
            	\xlabel{GTA-IM \cite{ref:CG20}},
            	\xlabel{HumanEva \cite{ref:SB09}},
            	\xlabel{Ikea FA \cite{ref:TC17}},
            	\xlabel{Kinder-Gator 2.0 \cite{ref:DA20}},
            	\xlabel{LaFAN1 \cite{ref:HY20}},
            	\xlabel{MADS \cite{ref:WZ17}},
            	\xlabel{MPI-INF-3DHP \cite{ref:MR17}},
            	\xlabel{MSR Action 3D \cite{ref:LZ10}},
            	\xlabel{Penn Action \cite{ref:ZZ13}},
            	\xlabel{PoseTrack \cite{ref:AI18}},
            	\xlabel{PROX \cite{ref:HC19}},
            	\xlabel{WBHM \cite{ref:MT15}}
            },
            xticklabel style={font=\tiny},
            %
            % Y axis, ticks & labels
            y=0.008\linewidth,
            ymin=0, ymax=45,
            ytick={0, 5, 10, 15, 20, 25, 30, 35, 40, 45},
            yticklabel style={font=\tiny},
            %
            % axis, ticks & grid
            axis line style={draw=none},
            major tick length=0,
            ymajorgrids,
            xmajorgrids=false,
        ]
            \addplot[
                white,
                fill=plotcolor1,
            ] coordinates {
                (1, 43)
                (2, 42)
                (3, 8)
                (4, 7)
                (5, 4)
                (6, 4)
                (7, 4)
                (8, 3)
                (9, 3)
                (10, 2)
                (11, 2)
                (12, 2)
                (13, 2)
                (14, 2)
                (15, 2)
                (16, 1)
                (17, 1)
                (18, 1)
                (19, 1)
                (20, 1)
                (21, 1)
                (22, 1)
                (23, 1)
                (24, 1)
                (25, 1)
                (26, 1)
                (27, 1)
                (28, 1)
                (29, 1)
                (30, 0)
            };
        \end{axis}
    \end{tikzpicture}
    \caption{
        Histogram of the number of papers using a given dataset among the works in skeleton-based deep human animation covered in this survey.
    }
    \label{fig:datasets}
\end{figure}
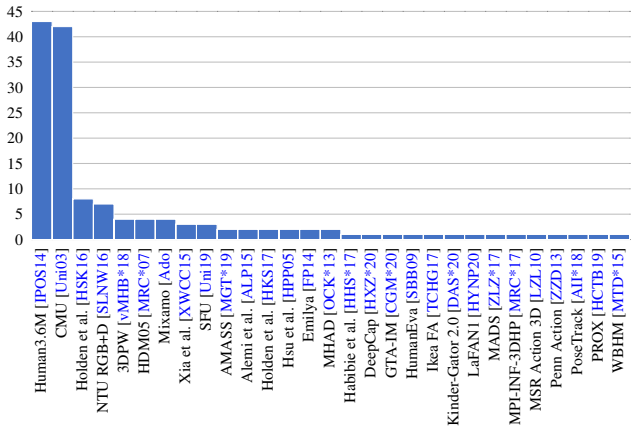

    Beyond these two standard human motion databases, a few others are noticeable, e.g. for the types of motion they contain, for the annotations that are included, or even for the environment in which they were captured. 
    % HDM05
    A few years after the release of CMU, M\"{u}ller~\etal~\cite{ref:MR07} proposed HDM05, a public well-documented database of systematically recorded motion capture data. Complementary to CMU which contains a large number of diverse motion sequences, HDM05 is composed of a limited number of specific motion sequences (one hundred) which were executed from 10 to 50 times by five actors. As an example, a cartwheel starting with the left hand has been performed 21 times.
    %
    % NTU RGB+D
    More recently, Shahroudy~\etal~\cite{ref:SL16} proposed NTU RGB+D, one of the largest datasets for 3D skeleton-based action recognition. It contains 60 action classes as well as RGB and infrared (IR) videos and depth map sequences (D) synchronized with motion sequences. Later on, Liu~\etal~\cite{ref:LS20} added another 60 classes to constitute NTU RGB+D 120, roughly doubling the size of the dataset. However, the motion sequences in these two datasets are represented only by joint positions which limits their use in skeleton-based human animation.
    %
    % 3DPW
    Since motion capture systems need dedicated environment \eg for a multicamera setup, human motion datasets are mostly captured in the lab, resulting in a lack of in the wild data. To this end, von Marcard~\etal~\cite{ref:VH18} proposed a pose estimation method, leveraging inertial measurement units in addition to a hand-held camera, accurate enough to capture a new dataset called 3DPW consisting of human pose sequences in the wild synchronized with RGB videos. It contains challenging sequences including walking in the city, going up-stairs, or taking the bus for a total of more than 51000 frames.

    % Mixamo
    A special need of data occurs from the use-case of retargeting (see Section~\ref{subsection:Retargeting}), which consists in transferring the movements of a character to another one with a different morphology, \ie motion data with diversity among morphologies. Unfortunately, most human motion datasets feature only a very limited number of subjects with minor morphological differences. As a result the so-called Mixamo Dataset~\cite{ref:MI21} is often used to train or evaluate models for retargeting. Indeed Mixamo is a computer graphics technology company developing services for 3D character animation including an online animation store with downloadable animation sequences performed by numerous 3D character models with varied morphologies. This diversity of character models is crucial to the task of retargeting.

    % HS16
    Finally, existing datasets were also gathered to build larger human motion databases. Holden~\etal~\cite{ref:HS16} constructed a dataset by collecting CMU~\cite{ref:CM03}, HDM05~\cite{ref:MR07}, MHAD~\cite{ref:OC13} and Xia~\etal~\cite{ref:XW15} in addition to internal motion capture sequences. This data was retargeted to a uniform skeleton structure and resampled to 120 frames per second.
    %
    % AMASS
    More recently, Mahmood~\etal~\cite{ref:MG19} proposed AMASS, which unifies $15$ different optical marker-based motion capture datasets (including CMU~\cite{ref:CM03}, HDM05~\cite{ref:MR07}, SFU~\cite{ref:SF19}, HumanEva~\cite{ref:SB09}). The size of AMASS, initially around 42 hours of data, is still increasing. Motion sequences in AMASS are parameterised using the Skinned Multi-Person Linear (SMPL) model~\cite{ref:LM15}, a learnt model of human body shape and pose that provides a parameter space from which the skeleton, the joint orientations and the body mesh can be computed.

\subsection{Learning Spatial Features}
\label{subsection:SpatialFeatures}
    Besides the pose representation and the dataset, the network architecture can also have a significant impact on the deep representation learned. In this section, we present different types of architectures leveraged to learn spatial correlations in motion data. While most \acrShortPlr{DNN} designed to address tasks related to skeletal character animation do not exploit the prior knowledge we have about geometric and structural aspects of the human skeleton, \eg its symmetry or its hierarchical structure, a few methods proposed various clever architectures to benefit from this prior knowledge. We present them in the following sub-sections, divided into three categories: spatially-structured architectures, \acrFullPlr{CNN} and \acrFullPlr{GCN}.
    
    \subsubsection{Spatially-Structured Architectures}
        A first group of approaches to help \acrShortPlr{DNN} learn spatial correlations rely on network architectures structured around the human skeleton, such that the function computed by the network intrinsically encodes human skeleton characteristics. These approaches split the skeleton into body parts and process the corresponding data either in parallel network branches or hierarchically. 
        
        In parallel approaches, the main difference is usually related to the targeted task, which conditions the architecture of individual branches. Wang and Neff~\cite{ref:WN15} extracted deep motion signatures with an independent autoencoder for each branch (\ie limb or torso) and concatenated the outputs. Guo and Choi~\cite{ref:GC19} relied instead on fully connected layers for each branch, whose outputs are merged using a shared layer to predict the next frame. Nakada~\etal~\cite{ref:NZ18} similarly divided the body into separate modules responsible for controlling muscle activations of body parts from preprocessed common visual information. Finally, Jain~\etal~\cite{ref:JZ16} predicted future poses with one \acrFull{RNN} per body part, whose inputs are the neighbouring \acrShortPlr{RNN} predictions at the previous timestamps as well as their own previous predictions.
        
        Hierarchical approaches can be either top-down or bottom-up. Wang~\etal~\cite{ref:WH19} proposed a spatial encoder that separates the human pose into five body parts, then encodes and merges them two by two recursively. Both Li~\etal~\cite{ref:LiW19} and B{\"u}tepage~\etal~\cite{ref:BB17} used a similar top-down approach with a finer human pose split at the input. In contrast, Aksan~\etal~\cite{ref:AK19} proposed a bottom-up scheme where the human pose is predicted step by step from the root joint (\eg pelvis) to the end effectors, \ie the root is first predicted and then the other joints are recursively predicted using neighbouring predictions as additional inputs.

    \subsubsection{Convolutional Neural Networks}
    \label{subsection:SpatialFeaturesCNN}
        Another type of architecture sometimes employed to model spatial correlations relies on \DD convolutions, \ie in the spatial and temporal domains. To this purpose, the skeleton graph is flattened along the spatial dimension. \acrShortPlr{CNN} are particularly efficient at learning spatial correlations in data whose structure is regular such as images. However, learning the spatio-temporal dynamics of human joints remains challenging with \acrShortPlr{CNN} because the graph structure of the human skeleton cannot be meaningfully flattened along a single dimension. To capture the spatial correlations of joints from different limbs, Li~\etal~\cite{ref:LZ18} proposed to enlarge the convolutional kernels in the spatial domain. More recently, Zang~\etal~\cite{ref:ZP20} proposed to adaptively model the spatial correlations with deformable convolutional kernels whose relative positions of the entries are learned. The problem of learning patterns in irregular data structures like human poses with \acrShortPlr{CNN} can be overcome by extending convolutions to graph-structured data, as we will see next.

    \subsubsection{Graph Convolutional Networks}
    \label{subsection:SpatialFeaturesGNN}
        To leverage \acrShortPlr{CNN} while properly handling the graph-structure typically used in animation to represent skeletons, \acrFullPlr{GCN}, an extension of \acrShortPlr{CNN}, have been recently considered in different frameworks working on human motion data. \acrShortPlr{GCN} come in two different flavours~\cite{ref:BrB17}. Spatial approaches map neighbourhoods of each node in the graph to Euclidean patches on which a convolution is applied. Spectral approaches operate in the Fourier domain of the feature signals sampled on the graph, which depends on the graph Laplacian operator~\cite{ref:SN13}. By analogy with the convolution theorem, convolution filters are defined as spectral coefficients that are multiplied by the Fourier transforms of the signals.
        
        Aberman~\etal~\cite{ref:AbL20} resorted to a simple implementation of spatial \acrShortPlr{GCN}. The supports of convolution kernels around each joint are defined as $d$-ring neighbourhoods on the skeleton graph in the spatial dimension and extended to the temporal axis to obtain \DD skeleto-temporal convolutions. The operation of such \acrShortPlr{GCN} is limited to skeletons sharing the same topology. However, the motion retargeting network they proposed includes skeletal pooling/unpooling layers, based on the fusion/duplication of the signals of adjacent edges. The pooling layers bring the input topology to a common primal skeleton on which the core processing is performed. The result is transformed back to the original topology by the unpooling layers. Such a network can cope with any topology that is homeomorphic to the primal skeleton.
        
        The other contributions leveraging \acrShortPlr{GCN}~\cite{ref:ML19, ref:CS20, ref:LC20, ref:LC21} relied on the spectral approach of Kipf and Welling~\cite{ref:KW17}. Here, the output $F^{l+1}$ of the graph convolutional layer $l$ fed with a feature signal $F^{l}$ is formulated as $F^{l+1} = \sigma(\hat{A} F^{l} W^{l})$ where $W^{l}$ is the tensor of learnable convolution filter weights, $\hat{A}$ depends only on the graph adjacency matrix and $\sigma$ is a non-linear activation function. In all of the proposed approaches the weights of the adjacency matrix are learnt in addition to the convolution filter. Mao~\etal~\cite{ref:ML19} built their adjacency matrix from a fully connected graph of joints and thereby simultaneously learnt the motion correlations between joints that are physically connected and joints that are far apart but whose motion are dependent, \eg hands and feet during walking. Cui~\etal~\cite{ref:CS20} objected that this scheme may result in unstable training and separately learnt two graph adjacency matrices, one in which the weights of non-connected joints in the skeleton are forced to 0 and another with full joint connectivity. Two papers by Li~\etal~\cite{ref:LC20, ref:LC21} proposed \acrShortPlr{GCN} that operate at multiple scales, the finest scale corresponding to individual joints and the coarser scales to increasingly large body parts. While scales are analysed in parallel branches in~\cite{ref:LC21}, in~\cite{ref:LC20} the features at various scales are additionally fused within each graph convolutional block.

\subsection{Learning Temporal Features}
\label{subsection:TemporalFeatures}
    The temporal dimension of motion data is informative of the nature of the action being performed as well as the way it is performed. In the following sub-sections we review the approaches taken to model human dynamics, most of which rely either on \acrShortPlr{RNN} or \acrShortPlr{CNN}.
    
    \subsubsection{Recurrent Neural Networks}
        \acrShortPlr{RNN} are neural networks designed to process each timestep of a time series one after another, and can thereby handle variable length sequences. They maintain an internal state that captures the temporal context of the signal. \acrShortPlr{RNN} are most of the time based on \acrFull{LSTM} or \acrFull{GRU}.

        \paragraph*{Long Short-Term Memory.}
            A common \acrShort{LSTM} is composed of a memory cell that remembers values over arbitrary time intervals, and three gates -- an input gate, an output gate and a forget gate -- to regulate the flow of information into and out of the cell and to avoid a common problem with \acrShortPlr{RNN} known as the vanishing (or exploding) gradient problem. In general, the problem is that the gradients used to update the network weights can become extremely small (or large), either preventing the network from further learning or making the network diverge, respectively. In the case of \acrShortPlr{RNN}, the backpropagation through time heavily relies on the chain rule to compute gradients which exponentially decrease (vanishing problem) or increase (exploding problem) if any weight is greater or smaller than 1, respectively.%
            
            The memory cell remembers values over arbitrary time intervals, making \acrShort{LSTM} effective at capturing both short-term and long-term temporal dependencies. Indeed, \acrShortPlr{LSTM} have proven to be powerful for learning temporal dependencies by achieving state-of-the-art performance in key applications \eg \acrShort{NLP}, machine translation, etc. In human motion related problems, \acrShortPlr{LSTM} have been broadly employed, \eg in motion prediction~\cite{ref:FL15, ref:CA19, ref:LiuW19, ref:KG19, ref:JZ16} and generation~\cite{ref:HA20, ref:HY20, ref:GS17, ref:WY20, ref:WH19, ref:WA20}, in both physical~\cite{ref:WM17, ref:HT17, ref:MA19, ref:MT20} and kinematic~\cite{ref:WC17, ref:LL18, ref:WC21} character control, as well as in motion cleaning~\cite{ref:ZL18} and style transfer~\cite{ref:WC18}.
        
        \paragraph*{Gated Recurrent Unit.}
            As an alternative to \acrShortPlr{LSTM}, \acrShortPlr{GRU} have also been widely used, such as in motion prediction~\cite{ref:MB17, ref:TC17, ref:GW18a, ref:GW18b, ref:PG18, ref:GM19, ref:WA19, ref:XL19, ref:GC19, ref:CP20, ref:AA20, ref:LC20}, in motion generation~\cite{ref:YK20, ref:BK18, ref:AS20, ref:GW20} or in motion editing~\cite{ref:VY18, ref:JL20}. \acrShortPlr{GRU} rely on a gating mechanism similar to \acrShortPlr{LSTM} in order to avoid the vanishing gradient problem but have only two gates, a reset gate and an update gate. As a result, \acrShortPlr{GRU} use fewer parameters and therefore less memory, are computationally less expensive~\cite{ref:GC19} and thus train faster than \acrShortPlr{LSTM}. They can process entire motion datasets~\cite{ref:MB17} instead of training action-specific models~\cite{ref:FL15, ref:JZ16}. However, as shown by Weiss~\etal~\cite{ref:WG18}, the \acrShort{LSTM} is strictly stronger than the \acrShort{GRU} as it can easily perform unbounded counting, while the \acrShort{GRU} cannot. Thus, \acrShortPlr{LSTM} seem more accurate than \acrShortPlr{GRU} on longer sequences. In summary, the choice between \acrShort{LSTM} and \acrShort{GRU} depends on the processed data and the considered application.
        
        % Misc
        Besides pure \acrShort{LSTM} or \acrShort{GRU}, extensions~\cite{ref:TM18} or combinations of both~\cite{ref:WA19} have been used to model human dynamics. \acrFullPlr{BiLSTM} stack two \acrShortPlr{LSTM} running forward and backward, respectively. As a result, temporal information is processed and preserved in both directions, \ie past and future, which is helpful on certain tasks. For instance, \acrShortPlr{BiLSTM} are leveraged to map an example motion into an embedding within an imitation learning framework~\cite{ref:WM17}, to build a long plausible motion sequence from a set of short motion clips~\cite{ref:XX20} or even to refine \DDD motion data~\cite{ref:LZ19, ref:LZ20}.
        
    \subsubsection{Temporal Convolutions}
    \label{subsubsection:TemporalCNN}
        \acrShortPlr{CNN} provide an alternative to \acrShortPlr{RNN} for learning temporal patterns in motion data. They can be either \D along the temporal dimension, or use \DD spatio-temporal convolutions. Stacking several convolutional layers can efficiently capture both short and long range temporal patterns since lower and higher layers will capture dependencies between nearby and distant frames, respectively. \acrShort{CNN}-based approaches are more computationally efficient than \acrShort{RNN}-based ones because they process whole motion segments at once rather than frame by frame, allowing greater parallelism. However, \acrShort{CNN}-based architectures often contain elements that do not allow variable length inputs (\eg a few fully connected layers after convolutions) limiting their use in practice. As a result, they are quite rare in motion synthesis~\cite{ref:HS16, ref:HG19, ref:DH19}, prediction~\cite{ref:BB17, ref:PF20, ref:CS20, ref:CS21} or character control~\cite{ref:HB20} with regard to \acrShortPlr{RNN} but more frequent in other tasks where fixed-length motion sequences are more suitable, \eg motion editing~\cite{ref:LimC19, ref:ZY20, ref:SG20, ref:SA20, ref:KP20, ref:AbL20, ref:HoH17, ref:AW20, ref:DA20, ref:LA21} (see Section~\ref{section:MotionEditing}).

    \subsubsection{Miscellaneous}
        Other approaches to better model the temporal flow of human motions include motion phase representation, spectral decomposition of motion data and spatio-temporal attention. 
        % Phase
        For instance, several authors investigated the representation and learning of the phase of movements in the context of kinematic character control, which is detailed in section~\ref{subsection:KinCharCtrl}. In the character controller network proposed by Holden~\etal~\cite{ref:HK17}, the network weights are computed as a spline function of the phase, whose control points, representing network weights configurations during the human locomotion cycle, are learned. Other works~\cite{ref:ZS18, ref:SZ19, ref:SZ20, ref:LingZ20} use a gating network instead of the cyclic phase to blend expert weights, resulting in a mixture-of-experts scheme.  
        % Discrete Cosine Transform
        In the context of motion prediction, Mao~\etal~\cite{ref:ML19, ref:ML20} and Cai~\etal~\cite{ref:CH20} learned temporal correlations in the frequency domain by applying at the input and the output of their network a \acrFull{DCT} and its inverse, respectively.
    \begin{figure*}[t]
	\centering
	\mbox{} \hfill
	\includegraphics[width=0.75\linewidth]{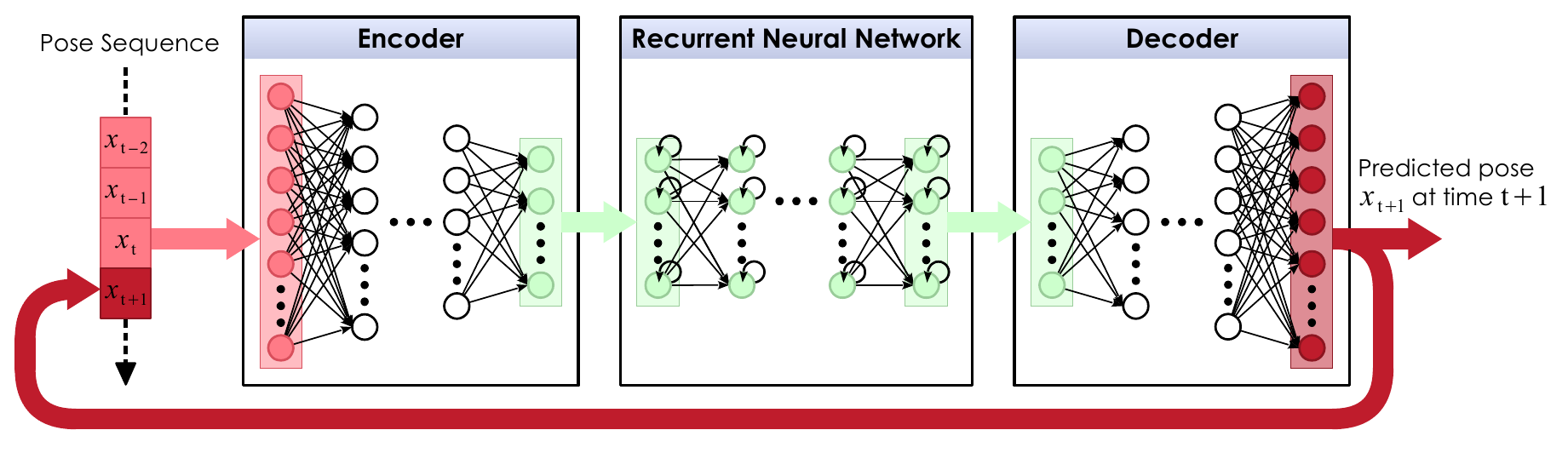}
	\hfill \mbox{}
    \caption{
        Illustration of the \acrFull{ERD} architecture originally proposed by Fragkiadaki~\etal~\cite{ref:FL15} and leveraged in both short and long term motion prediction in its original formulation and modified/extended versions. An \acrShort{RNN} captures motion dynamics in a latent space. The encoder and decoder feedforward \acrShortPlr{DNN} map skeletal poses to this latent representation and back.
    }
    \label{fig:ERD}
\end{figure*}

\section{Motion Synthesis}
\label{section:MotionSynthesis}
    Synthesising motion data is of strong practical interest to the media and entertainment industry. Besides generating realistic and diverse sequences, a key challenge is to be able to control various aspects of the motion using high-level parameters. In this section we distinguish the tasks of short-term (Section~\ref{subsection:ShortTermPrediction}) and long-term (Section~\ref{subsection:LongTermPrediction}) motion prediction, and motion generation (Section~\ref{subsection:DeepGenerativeModeling}). The first focuses on the short-term deterministic extrapolation of motion, \ie up to one or two seconds, from a past conditioning clip, while the second operates beyond this temporal horizon, with the purpose of generating plausible and diverse continuation of the observed motion. Indeed, reproducing ground-truth animations quickly becomes impossible because of the stochastic nature of human motion~\cite{ref:FL15, ref:PG18, ref:ZL18}. On the contrary, the ability to generate diverse sequences is often a desired feature in long-term motion prediction. In the third category, the goal is to synthesise from non-historical parameters pose sequences that are both consistent with the distribution of samples in a reference training dataset, and sufficiently diverse to capture its variations. We restrict our survey to general-purpose approaches and exclude works targeting application-specific contexts such as the synthesis of gestures from speech or dance animations from music. We summarise the methods presented in this section in Table~\ref{tab:synthesis}.

\subsection{Short-term Prediction}
\label{subsection:ShortTermPrediction}
    Short-term motion prediction consists in observing the motion on a given temporal horizon and predicting the motion for the near future. Let us note $X_t$ the \DDD pose of a skeleton at time~$t$: when observing the set of poses $X_{t-T_p},\dots,X_t$, we aim at predicting future poses $Y_{t+1},\dots,Y_{t+T_f}$ where $T_p$ and $T_f$ denote the past and future temporal observation and prediction windows, respectively.
    Motion prediction is useful for many applications including robotic navigation in crowds, human-robot interaction, video surveillance, virtual reality experiences or cloud gaming. In particular, the capability to predict the \DDD positions of human end effectors rather than trajectory provides a much richer and helpful information.

    Classical approaches have been developed roughly between $2000$ and $2015$ using techniques such as spatio-temporal autoregressive models, Hidden Markov Models or Gaussian processes. These \enquote{classical} techniques are out of the scope of this section. With the development of \acrShort{DL} techniques, as well as the production of larger mocap databases (as presented in section~\ref{subsection:Datasets}), it has become a natural path to use \acrShort{DL} for human motion prediction. There are multiple difficulties in this particular task of predicting human motion with \acrShort{DL}, namely:
    \begin{itemize}
        \item When the motion is stationary and periodic, the repeating patterns can be discovered and used for future prediction. Unfortunately, human motion in real-world scenario is rarely stationary. The first challenge is thus to design deep models that account for aperiodic and non-stationary motion.
        
        \item Articulated human motion can only be efficiently predicted if spatial and temporal dependencies can be captured. For instance, modelling the dependency between the right arm and the left leg during a walk cycle. The second challenge is to model these spatial and temporal dependencies.
    
        \item \acrShort{DL} offers the advantage of leveraging large quantities of data. However, the plausibility of the future prediction given the past observations should be explicitly enforced. In particular, Martinez~\etal~\cite{ref:MB17} showed that many classical \acrShort{ML} approaches can be beaten by a simple baseline: the zero-motion prediction (this can partially be explained by the discontinuity of the first frame prediction). The third challenge is therefore to ensure the plausibility and realism of the predicted motion. 
   
        % trick to avoid splitting next item over two columns 
        \end{itemize} \pagebreak \begin{itemize} 
        
        \item Humans rarely act in an empty environment. Constraints such as objects, or other humans in the scene, should be taken into account for prediction. The fourth challenge is to predict future motion in a dynamic and constrained environment.
    \end{itemize}
    We first review the existing methods, organised w.r.t. framework used. Then, we examine the existing answers that authors have proposed to the four aforementioned challenges.

\subsubsection{Classes of methods} \label{subsection:STP-methods}
    Various approaches have been applied to the problem of short-term motion prediction, and are presented according to the framework they built on, namely: \acrFull{Seq2Seq}, direct feedforward networks, \acrFullPlr{GCN} and \acrFullPlr{GAN}.

    \paragraph*{Sequence to Sequence.}
    Popular and widely used in motion prediction, \acrShort{Seq2Seq}-based approaches generally consist in training an \acrShort{RNN} in some latent space. Both the \acrShort{RNN} and the hidden latent representation are trained jointly. Known issues of these techniques are the convergence to a mean pose, as well as finding the trade-off in exposing the model to its own errors so that it can recover from deviations. We describe hereafter the main solutions proposed in the literature.
    
    Fragkiadaki~\etal\cite{ref:FL15} pioneered the \acrFull{ERD} architecture, in which an \acrShort{LSTM} network is trained in the hidden layer of an autoencoder, enabling the network to learn a representation suited for motion prediction, as illustrated in Figure~\ref{fig:ERD}. Papers that pursued this scheme generally opted for a \acrShort{GRU} module afterwards.
    
    Some papers incorporated the skeletal structure in the network. Jain~\etal~\cite{ref:JZ16} learnt motion dynamics using a spatio-temporal graph modelling body parts, using an \acrShort{RNN} architecture, which can be viewed as a structural-\acrShort{RNN} architecture. Aksan~\etal~\cite{ref:AK19} proposed a structured prediction mechanism, composed of a hierarchy of sub-layers connected to the kinematic chain of the skeleton. This enables an explicit modelling of body part motion prediction at a local and global scale. This structured prediction is compatible with different flavours of joint representation (exponential map or unit quaternions), as well as different types of prediction networks (simple \acrShort{RNN}, \acrShort{Seq2Seq}, etc.). Li~\etal~\cite{ref:LiW19} proposed an approach where the \acrShort{RNN} is trained on a hierarchical decomposition of joints, as well as on two different temporal horizons to maintain global consistency. Yang~\etal~\cite{ref:YK21} used an \acrShort{RNN} network based on \acrShortPlr{GRU} to predict the lower body motion, given past observations of the upper body.
    
    Although the prediction of joint positions is usually preferred, some authors opted for different pose representations and the associated loss function. To improve recent \acrShort{RNN} approaches, Martinez~\etal\cite{ref:MB17} proposed the following improvements: first, instead of feeding unrealistic noise at learning, they incorporated network predictions to drive the network to recover from its own mistakes. Second, they proposed a residual architecture that takes into account first-order derivatives. Hence, they used a \acrShort{GRU} in the latent representation, forcing weight sharing between encoder and decoder: motion continuity is enforced through residual connections. Chiu~\etal~\cite{ref:CA19} learned a hierarchical and multi-scale latent representation where an \acrShort{RNN} is trained to predict human velocities. Gui~\etal~\cite{ref:GW18a} proposed a \acrShort{GRU}-based \acrShort{Seq2Seq} architecture, incorporating adversarial training to decrease the short-term discontinuity and increase the long-range realism. In addition, they advocated that the loss function should be based on a geodesic loss, leveraging the properties of the Lie structure of orthogonal matrices. Pavllo~\etal~\cite{ref:PG18,ref:PF20} proposed an autoregressive model based on two \acrShort{GRU} layers in the space of quaternions, therefore avoiding the gimbal lock issue at the expense of constraining quaternions to remain on the unit 4-sphere (see Section~\ref{subsection:AngPoseRepr}). The total loss for the pose is composed of a quaternion loss and a positional loss obtained after \acrShort{FK}.
    
    Finally, some methods took inspiration from the \acrShort{Seq2Seq} framework but replaced the \acrShort{RNN} prediction layer: Li~\etal~\cite{ref:LZ18} used two encoders to learn latent representations of both short and long term horizons. A convolutional architecture is trained instead of an \acrShort{RNN} to predict future poses.  Xu~\etal~\cite{ref:XL19} proposed a hierarchical \acrShort{Seq2Seq} model that encodes each sequence into a latent representation. The prediction is then seen as a completion in this latent representation, obtained by vector addition. A single fully connected completion network is trained to this goal. Gui~\etal~\cite{ref:GW18b} leveraged a meta-learning scheme within a \acrShort{Seq2Seq} approach using \acrShortPlr{GRU} to adapt the model's parameters to new training examples. 
    To the best of our knowledge, the work of Wang~\etal~\cite{ref:WA19} is the only approach using \acrShort{DRL} for motion prediction. The prediction problem is decomposed into $K$ steps, and the progressive prediction is trained using an imitation learning approach.
    
    \paragraph*{Direct Feedforward.}
    Less used than \acrShort{Seq2Seq}-based methods, direct feedforward techniques aim at computing the prediction network $\mathcal{F}$ directly from the entire set of poses given the past observations: $Y_{t+1},\dots,Y_{t+T_f}= \mathcal{F} (X_{t-T_p},\dots,X_t)$. 
    
    B{\"u}tepage~\etal~\cite{ref:BB17} trained such a feedforward network after a temporal encoding of human motion. Feature learning is evaluated thanks to an action classifier. Guo and Choi~\cite{ref:GC19} proposed a combination of two \acrShortPlr{DNN} for human motion prediction: the first network, called \emph{SkelNet}, assembles the prediction of different body parts, while the second network, called \emph{Skel-TNet}, accounts for temporal dependencies modelled through an \acrShort{RNN}. Zang~\etal~\cite{ref:ZP20} proposed a deformable spatio-temporal convolution approach that captures in the past sequence the most relevant poses, through a masking mechanism. Based on this temporal attention, parameters for the prediction of the future motion are generated on the fly, as in few-shot learning methods. Cai~\etal~\cite{ref:CH20} leveraged the increasingly popular transformer architecture~\cite{ref:KN21}. Originally designed for processing data sequences in \acrShort{NLP}, it relies on attention mechanisms to better account for long-range dependencies than \acrShortPlr{RNN}. The attention weights learnt in a transformer model capture the relevance of data sequence items to other items that can be far apart. Because these weights modulate the input data prior to convolution, transformer networks can be viewed as extensions of \acrShortPlr{CNN} and \acrShortPlr{RNN} in which the learnt convolution kernels are made data-dependent. This increased expressiveness comes at the price of larger training data volume requirements. More specifically, in~\cite{ref:CH20} joints are converted using a \acrFull{DCT}. The decoding of the future pose is done progressively in accordance with the skeleton topology.

    \paragraph*{Graph Convolutional Networks.}
    As described in Section~\ref{subsection:SpatialFeaturesGNN}, \acrFullPlr{GCN} leverage the graph nature of the human skeleton, and have therefore been used to design prediction methods that act on such graphs. They were first used in the context of short-term motion prediction by Mao~\etal~\cite{ref:ML19}, who proposed an approach where poses are encoded in a latent space using \acrShort{DCT} decomposition. This representation is then fed to a \acrShort{GCN} for feedforward prediction, while the connectivity of the graph is learned through the adjacency matrix, which enables to learn the temporal and spatial dependencies between body joints. They also proposed an attention mechanism applied to the \acrShort{DCT} coefficients of the observed sequence~\cite{ref:ML20}, where the main idea is to focus on relevant history information to predict the future poses. Approaches based on \acrShortPlr{GCN} have also been proposed to capture skeletal and motion information at different scales. For instance, Li~\etal~\cite{ref:LC20,ref:LC21} used \acrShortPlr{GCN} to capture the skeleton structure over different scales, ranging from individual joints to increasingly large body parts. A cross-scale fusion block then merges the features over scales and feeds this representation into a \acrShort{GRU}-based decoder~\cite{ref:LC20}. As an alternative, scales can be analysed in parallel branches to extract features used both for action recognition and motion prediction~\cite{ref:LC21}. The predicted motion category is then used in the motion prediction network. Similarly, in order to capture relevant motion information at different scales, Lebailly~\etal~\cite{ref:LK20} proposed a temporal inception module, using \D temporal convolutions at different scales, combined with a \acrShort{GCN} to predict the future poses. Unlike previous approaches that represent skeletal connections with a single adjacency matrix, Cui~\etal~\cite{ref:CS20} relied on a double graph convolutional approach to better learn the dynamic relationships between skeletal joints, therefore learning a connective graph to represent the natural kinematic links of the human skeleton and a global graph to account for the dynamics of non-connected joints. Convolutions over these two graphs are then used to predict future poses. \acrShortPlr{GCN} have also been used to explore the prediction of human-object interactions: Corona~\etal~\cite{ref:CP20} built on the work of Martinez~\etal~\cite{ref:MB17} and explicitly incorporated the modelling of such interactions. As adjacency matrices typically vary in time in these dynamic interactions, the resulting graph representation is learned to capture their dynamics.

\subsubsection{Spatial and Temporal Dependencies}
    One difficulty in motion prediction is to leverage the spatial and temporal dependencies that exist in human motion. In the spatial domain, human joints are directly related to their parents but indirectly related to other joints (\eg symmetry or regularity in motion). In the temporal domain, the repeating patterns (or information redundancy) can greatly help in predicting the future motion. Modeling the motion over various scales greatly helps for short-term motion prediction. This is closely related to attention mechanisms that have been well studied in the \acrShort{DL} community. Researchers started to investigate spatial dependencies first, then rapidly studied both spatial and temporal dependencies especially through temporal attention mechanisms. We here review the different solutions that have been proposed in the literature.

    Jain~\etal~\cite{ref:JZ16} used a spatio-temporal graph that explicitly models the body structure (spine, arm, and leg) and captures the spatial dependencies. B{\"u}tepage~\etal~\cite{ref:BB17} learned a temporal encoding of motion using different time scales and convolutions, while a graph network is used to encode the hierarchical structure of the skeleton. Li~\etal~\cite{ref:LZ18} explicitly used two latent representations for short-range and long-range observation. By doing so, they expected to benefit from long-range motion consistency, as well as to improve the dynamics of short-term prediction. Cui~\etal~\cite{ref:CS21} used a temporal attention mechanism, jointly with dilated convolutions, to capture long-range motion features. Capturing temporal and spatial dependencies between body joints has also been explored using \acrShortPlr{GCN}~\cite{ref:ML19,ref:CS20,ref:LC20,ref:ML20,ref:LK20}, as already detailed in Section~\ref{subsection:STP-methods}, to match the graph nature of the human skeleton. Cai~\etal~\cite{ref:CH20} leveraged the transformers concept that learns the spatial correlations as well as the temporal smoothness of the predicted skeletons.

\subsubsection{Plausibility and Realism}
    Data-driven approaches to predict future motion can face the issue of predicting unrealistic poses in terms of biomechanics. In addition, motion prediction can suffer from a lack of realism whatever the temporal range of prediction. We present here the proposed solutions to these problems.

    To improve plausibility, Gui~\etal~\cite{ref:GW18a} included two discriminators at training time: a fidelity and a continuity discriminator. The fidelity discriminator assesses whether the prediction is smooth enough, while the continuity discriminator encourages the prediction to be consistent with past observations, thus limiting the discontinuities. It is also possible to rely on a discriminator to discriminate from real sequences, however results show that this improves the prediction by a small margin only~\cite{ref:LZ18}. Cui~\etal~\cite{ref:CS20} used the Gram matrix in the loss function to add more consistency between the predicted and ground-truth poses. They also included bone length preservation.
    
    Pavllo~\etal~\cite{ref:PG18} proposed to mix a \emph{teacher-forcing} approach (\ie feeding the network with ground-truth motions) with an approach where the network is fed with its own predictions. The network is first trained with teacher-forcing and progressively switches to its own predictions through a curriculum schedule. This progressive training improves the error and the model stability. Cai~\etal~\cite{ref:CH20} used a dictionary mechanism, similar to a memory cell. This dictionary encodes motion that could be similar but seen in slightly different contexts. Bourached~\etal~\cite{ref:BG21} proposed an out-of-distribution approach, where samples can be augmented thanks to a \acrFull{VAE} learned on Human3.6~\cite{ref:IP14} and CMU~\cite{ref:CM03} datasets. Using graph convolutional layers, the \acrShort{VAE} models connectivity, positions and temporal frequencies, which helps to limit the distribution shift and to regularise the training. More specifically, this mechanism helps producing larger quantities of plausible training data.

\subsubsection{Context, Environment and Interactions}
    Modelling human motion in an empty environment, without objects or humans, is already a difficult task. However, context, environment and surrounding characters are crucial for predicting motions in real-life scenarios, especially to avoid unrealistic situations, \eg character collisions or incorrect interactions with the environment. This is an even more complex task, mainly due to the lack of databases and the large variety of situations that can be encountered.
    
    To better predict interactions between a human and all objects of the scene, Corona~\etal~\cite{ref:CP20} proposed to incorporate the context and the interaction between humans and objects  in a novel context-aware motion prediction architecture. To do so, a convolutional method over the graph of objects and persons is learned, building on the \acrShort{RNN} implementation proposed by Martinez~\etal~\cite{ref:MB17}. The graph relationships are also learned, initialised so that the prediction of one object only depends on itself, and progressively accounting for interactions. Similarly, to account for both scene and social contexts in the prediction, Adeli~\etal~\cite{ref:AA20} proposed a \acrShort{Seq2Seq} approach with a \acrShort{GRU} architecture, where the scene and social contexts are captured independently through the analysis of monocular video. The spatio-temporal features are first extracted by a \acrShort{CNN}, then fed to the decoder module to account for context.

\subsection{Long-term Prediction}
\label{subsection:LongTermPrediction}
    The goal of long-term motion prediction is either to extrapolate a past conditioning clip to the future (\ie forecasting, see Section~\ref{subsection:Forecasting}) or to interpolate between known past and future character poses, often far apart both temporally and spatially (\ie in-betweening, see Section~\ref{subsection:Inbetweening}). Unlike short-term prediction approaches, the time horizons considered are larger and the synthesised motions are not constrained to reproduce the ground truth. Instead, the goal at large is to generate a diversity of plausible continuations of the original clip.

\subsubsection{Forecasting}
\label{subsection:Forecasting}
    Most of the approaches in this category are based on the \acrFull{ERD} model originally proposed by Fragkiadaki~\etal~\cite{ref:FL15} for motion prediction and presented in section~\ref{subsection:ShortTermPrediction}. The generation is conditioned by a past window of motion frames. The temporal context is captured by an \acrShort{RNN} operating in the latent space at the output of the encoder, as illustrated in Figure~\ref{fig:ERD}. As pointed out by many authors~\cite{ref:HH17, ref:ZL18, ref:LiuW19, ref:WC21, ref:GW20}, future motion predicted by \acrShort{ERD} networks tend to converge to a motionless state or to diverge from human motion. This is attributed to several issues that many of the approaches presented below aim at mitigating. Although some of these issues are also relevant to short-term motion prediction (section~\ref{subsection:ShortTermPrediction}), they are exacerbated when targeting longer prediction horizons.
    
    A major downside of the \acrShort{ERD} architecture is the accumulation of errors along time at the output of the \acrShort{RNN}. As a result, the quality of the generated motion degrades as the prediction moves away from the past conditioning sequence. Wang~\etal~\cite{ref:WC21} and Kundu~\etal~\cite{ref:KG19} treated the \acrShort{ERD} model as the generator of a \acrShort{GAN} and added a corresponding discriminator to improve the plausibility of the motion predictions. Kundu~\etal~\cite{ref:KG19} concatenated a stochastic component $r$ to the output of the \acrShort{ERD} encoder and complemented the discriminator with a critic network that regresses $r$ from the predicted motion sequence. The regressed value is fed back to the \acrShort{ERD} decoder to refine the predicted sequence. This encourages the learning of a one-to-one-mapping between the stochastic input to the \acrShort{GAN} and the generated motion, thereby minimising the risk that the output collapses to a single mode of the motion distribution, a well-known issue with \acrShortPlr{GAN}~\cite{ref:GP14}.

    Several authors~\cite{ref:HH17, ref:GW20} ascribed the regression towards a static pose to the ill-posedness of long-term motion prediction and advocated the use of external control signals to disambiguate the task. For instance, Ghorbani~\etal~\cite{ref:GW20} built the \acrShort{RNN} cell of their network around a conditional \acrShort{VAE} that is conditioned on external constraints, such as action type and character gender, as well as on the hidden state of the \acrShort{RNN} at the previous time step. Cao~\etal~\cite{ref:CG20} conditioned their motion prediction scheme on the environment the character moves in, specified using a \DD picture. The past motion history is provided as a sequence of \DD joint heatmaps in the scene. A first conditional \acrShort{VAE} network samples a plausible \DD target end location of the predicted motion sequence inside the environment, from which a second network predicts the \DDD trajectory of the character center. Finally, a third network promotes this point trajectory into a 3D pose sequence.

    The \acrShortPlr{RNN} at the core of the \acrShort{ERD} model create other issues. Training an \acrShort{RNN} using ground-truth future samples rather than exposing it to its own predictions creates a difference in behaviour between the training and inference stages that is detrimental to performance, a phenomenon referred to as exposure bias~\cite{ref:PG18, ref:GW20}. As a mitigation measure, Zhou~\etal~\cite{ref:ZL18} proposed an auto-conditioning network that is trained alternately in open loop on ground-truth motion and in closed loop on its own predictions. Gopalakrishnan~\etal~\cite{ref:GM19} strike a balance between the two training modes in their loss function and gradually increase the weight of the closed-loop term. \acrShortPlr{RNN} have a tendency to overly focus on recent poses and fail to capture long-term temporal dependencies of human motion. To counter this effect, Tang~\etal~\cite{ref:TM18} augmented their \acrShort{RNN} with a temporal attention mechanism that encourages pose predictions to align to previous poses in a pose embedding. To the same purpose, Liu~\etal~\cite{ref:LiuW19} proposed a hierarchical \acrShort{RNN} cell that captures a global motion context in addition to the \acrShort{RNN} hidden state at every frame. 

    Unlike for short-term motion prediction where the goal is to minimise deviations from the ground truth, diversity and randomness in the generated motion sequences is encouraged by complementing the past sequence input with a stochastic component. In \acrShort{ERD} networks this can be achieved by adding random noise to the latent pose representations at the output of the encoder~\cite{ref:XL19}, or by stacking a random vector to the latent representation of the \acrShort{RNN}~\cite{ref:KG19}. Stochasticity can also be embedded in the \acrShort{RNN} cell, as in~\cite{ref:GW20} where this cell is built around a conditional \acrShort{VAE} whose random latent code drives motion generation.

    Other works proposed more expressive models of the skeleton or of the motion dynamics to improve the realism of synthesised animations. For instance, representations of the articulated bone structure as quaternions~\cite{ref:PG18} or as \DDD rigid transformations in the Lie algebra $\se{3}$~\cite{ref:LiuW19} were shown to generate more natural motion sequences. Please refer to Section~\ref{subsection:PoseRepr} for a detailed presentation of these approaches. Gopalakrishnan~\etal ~\cite{ref:GM19} incorporated short-term motion history in their representation of pose by complementing joint angles with their local temporal derivatives up to order 3. The prediction of motion dynamics is cast by Wang~\etal~\cite{ref:WC21} in a probabilistic framework. The transition of character pose $x_t$ at time $t$ to the next frame is expressed by a probability density function $p(x_{t+1} \mid x_t)$, modelled as a multivariate \acrFull{GMM}. Their network builds on two \acrShort{RNN} cells that predict the \acrShort{GMM} parameters at each time step by maximising its likelihood.

    Many authors acknowledged that learning a motion manifold for a large diversity of styles and movements is a difficult problem. To facilitate this task, several authors~\cite{ref:GS17,ref:LiuW19,ref:GM19,ref:WH19} learned separate sub-networks for modelling the spatial and temporal components of human motion. The prediction network of Gopalakrishnan~\etal ~\cite{ref:GM19} is a two-level hierarchical \acrShort{RNN} where the upper layer sketches a trajectory for future motion while the lower layer synthesises poses on this trajectory. The decoder in the \acrShort{ERD} network architecture of Wang~\etal~\cite{ref:WH19} features a forward and a backward motion predictor, whose outputs are concatenated to reconstruct the full-length motion sequence. The authors argue that since the backward predictor forces the decoder to pay attention to the last few frames output by the encoder, it encourages the model to first learn short-term temporal correlations before accounting for longer-term dependencies.
    
    Unlike all previous approaches, Hernandez-Ruiz~\etal~\cite{ref:HG19} did not rely on an \acrShort{RNN}. They instead reformulated motion prediction as a temporal inpainting task and built their model on a \acrShort{GAN}. They fed the generator with the past conditioning motion clip, and encoded it into frame-specific latent representations that are processed at multiple scales before being decoded into an output pose sequence.

\subsubsection{In-betweening}
\label{subsection:Inbetweening}
    The task of interpolating motion between starting and ending character poses, often far apart, is referred to as \textit{In-betweening}. It can be viewed as a long-term prediction task conditioned on past and future contexts.

    Yu~\etal~\cite{ref:YK19} proposed a simple scheme for interpolating a motion sequence from starting and ending positions of end-effector joints. Their main objective is computational efficiency, in order to generate terrain-adaptive character motion in real time for video games. They cascaded two fully connected networks: the first interpolates the trajectories of the end effectors in time, the second infers full poses at each frame.

    The main purpose of other works is to generate plausible and diverse motion over long transitions bounded by a starting and an ending keyframe. Kaufmann~\etal~\cite{ref:KA20} leveraged a deep convolution denoising autoencoder, in which pooling layers ensure large receptive fields to capture long-range spatial and temporal joint correlations. A curriculum learning scheme feeds the encoder with sequences containing increasingly large temporal gaps to improve the training. Xu~\etal~\cite{ref:XX20} proposed a temporally hierarchical scheme in which the transition segment is split into equal length sub-segments. Trajectory constraints are provided at each sub-segment endpoint. The transition sequence is initialised by sampling a motion clip from the training dataset for each subsegment. Next, the style of each clip is changed to match the style of a reference sequence throughout the whole transition sequence. This stage leverages a motion autoencoder with separate content and style embeddings, that is trained in an unsupervised way. Style transfer is performed by linearly combining the content latent code of the initial clips and the style latent code of the reference style sequence. Finally, transitions between the endpoints of consecutive sub-segments are generated using forward and backward \acrShort{LSTM} networks, and the plausibility of the generated sequence is enhanced by combining the generation network with a discriminator in a \acrShort{GAN} framework. Harvey~\etal~\cite{ref:HY20} pointed out that in-betweening between distant keyframes may result in stalling or teleportation artifacts if the temporal evolution of the motion is not monitored during the generation process. To deal with this issue, they proposed an \acrShort{ERD} architecture that is fed with the pose representation deltas between the current and the target frame, in addition to the character pose at the current timestep and the end pose. The time-to-arrival is encoded in a sine wave and added, rather than stacked, to the latent code that is input to the \acrShort{RNN}, forcing the network to take this piece of information into account during training.

\subsection{Generative Synthesis} \label{subsection:DeepGenerativeModeling}
    Unlike motion prediction methods, approaches covered in this section synthesise human motion by processing a random seed with a deep generative network, optionally conditioned on semantic cues typically pertaining to the character trajectory or motion style. They are grouped in the following sub-sections based on the deep generative framework they build on. As an exception, in the scheme proposed by Holden~\etal~\cite{ref:HS16} the synthesis process is purely deterministic and driven by high-level cues.

\subsubsection{Restricted Boltzmann Machines}
\begin{figure}[t]
    \centering
    \includegraphics[width=\linewidth]{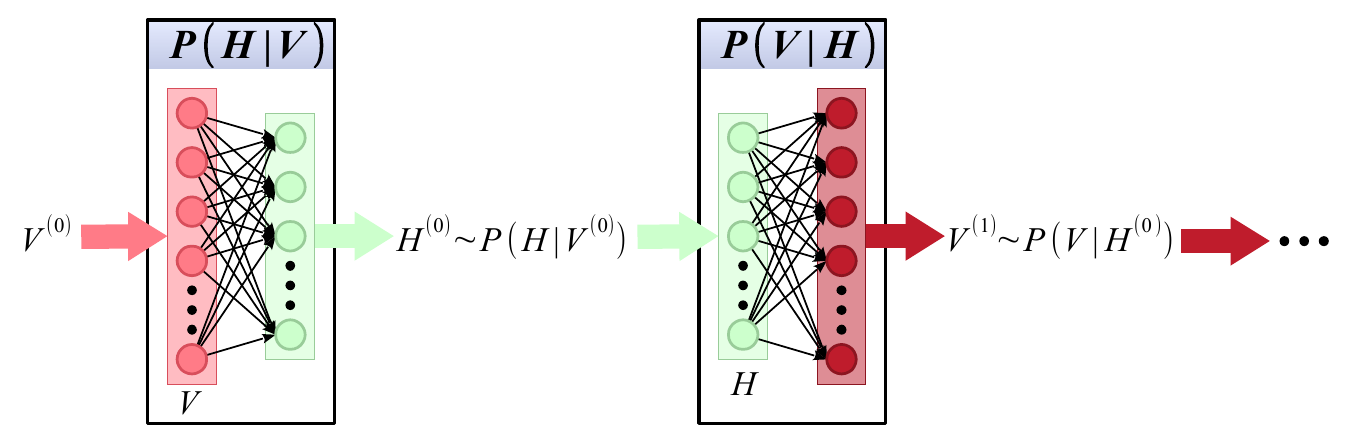}
    \caption{
        A \acrFull{RBM} can be represented by two fully-connected neural networks whose outputs provide the conditional probabilities of hidden variables $H$ given visible variables $V$ and vice-versa. Sampling from these probabilities builds a chain of hidden $H^{(j)}$ and visible $V^{(j)}$ samples that converges to the model distribution. The chain is initiated by drawing $V^{(0)}$ from the training set.
    }
    \label{fig:RBM}
\end{figure}

    Initially proposed by Smolensky~\cite{ref:S86}, \acrFullPlr{RBM} build on a parametric expression of the probability density function of the data distribution from which new samples are to be generated. This density depends on visible units $V$ that correspond to the variables to be synthesised and hidden units~$H$. An \acrShort{RBM} can be represented as a bipartite graph in which hidden units are conditionally independent given the visible units, and vice-versa. The form of the probability density function $P(H,V)$ in an \acrShort{RBM} provides simple closed-form expressions for $P(H|V)$ and $P(V|H)$ as fully connected layers with a sigmoid activation. These layers are usually trained using an approximate gradient ascent scheme known as \emph{Contrastive Divergence}~\cite{ref:H02}. Once $P(H|V)$ and $P(V|H)$ have been determined, new samples are generated using a \acrFull{MCMC} sampling process initiated by visible units $V^{(0)}$ drawn from the training dataset. Given $V^{(i)}$, $P(H|V^{(i)})$ is computed and hidden units $H^{(i)}$ are drawn from that distribution. Next, evaluating and sampling from $P(V|H^{(i)})$ yields $V^{(i+1)}$. This alternated sampling scheme between $V^{(i)}$ and $H^{(i)}$ is stopped after $K$ steps, $V^{(K)}$ providing the sought generated sample (see Figure~\ref{fig:RBM}).

    \paragraph*{Conditional RBMs.}
    Taylor~\etal~\cite{ref:TH06} extended the \acrShort{RBM} generative framework to the synthesis of time series by introducing an autoregressive model that conditions the visible and hidden variables at the current time step on the visible variables at past time steps. The formulation of this \acrFull{CRBM} differs from the original \acrShort{RBM} only by additional bias terms in the expressions of $P(H|V)$ and $P(V|H$). These biases are learnt during training. The optimisation and sampling processes are otherwise unchanged. Motion sequences can be generated using a \acrShort{CRBM} by mapping character joint positions or rotations to its visible units. When the training dataset contains different categories of motion, the generated category is dictated in principle by the selection of the initial \acrShort{MCMC} sample $V^{(0)}$, although transitions between different types of motion can be forced by adding noise to the hidden states.
    
    \paragraph*{Factored CRBMs.}
    The same authors~\cite{ref:TH09} proposed the \acrFull{FCRBM} architecture to better control the motion category or style. It is essentially a gated \acrShort{CRBM} consisting of three layers: the visible units at the previous time steps form the input layer, the same units at the current time step are the output layer, and the connections between these two layers are gated using multiplicative weights that form a third hidden layer. To obtain a more efficient representation, the third-order interaction tensor between the 3 layers is factored into three matrices of pairwise interactions. The weights in the hidden layer are defined as linear functions of a one-hot vector of motion style labels that control motion synthesis. Transitions between motion styles can be introduced at will by appropriate settings of the motion style labels over the sequence. Alemi~\etal~\cite{ref:AL15} applied the \acrShort{FCRBM} model to the generation of controlled affective variations of walking motion, based on time-varying valence and arousal labels.

    \paragraph*{Hierarchical FCRBMs.}
    Chiu and Marsella~\cite{ref:CM11} noted that \acrShort{CRBM} and \acrShort{FCRBM} approaches are prone to overfitting when trying to generalise to a rich set of styles and motion, because the training dataset can only sparsely sample the set of all possible style combinations and transitions. To mitigate this issue, they proposed a \acrFull{HFCRBM} model, corresponding to a \emph{Deep Belief Network} with an \acrShort{FCRBM} on top of a simplified \acrShort{CRBM} where the temporal dependency on past visible states has been removed. The dynamics of the synthesised sequence is controlled exclusively by the upper \acrShort{FCRBM}. In their multi-path approach, a separate \acrShort{FCRBM} is learnt for each motion style, the output visible layers are linearly combined with predefined weights defining the desired motion style, and the result is fed to the bottom layer of the \acrShort{HFCRBM}. Thus, rather than relying on a unique \acrShort{FCRBM} to represent all styles, the mixing of styles is performed by mixing individual \acrShort{FCRBM} instances within the hidden layer of the \acrShort{HFCRBM}.

\begin{figure}
    \centering
    \includegraphics[width=\linewidth]{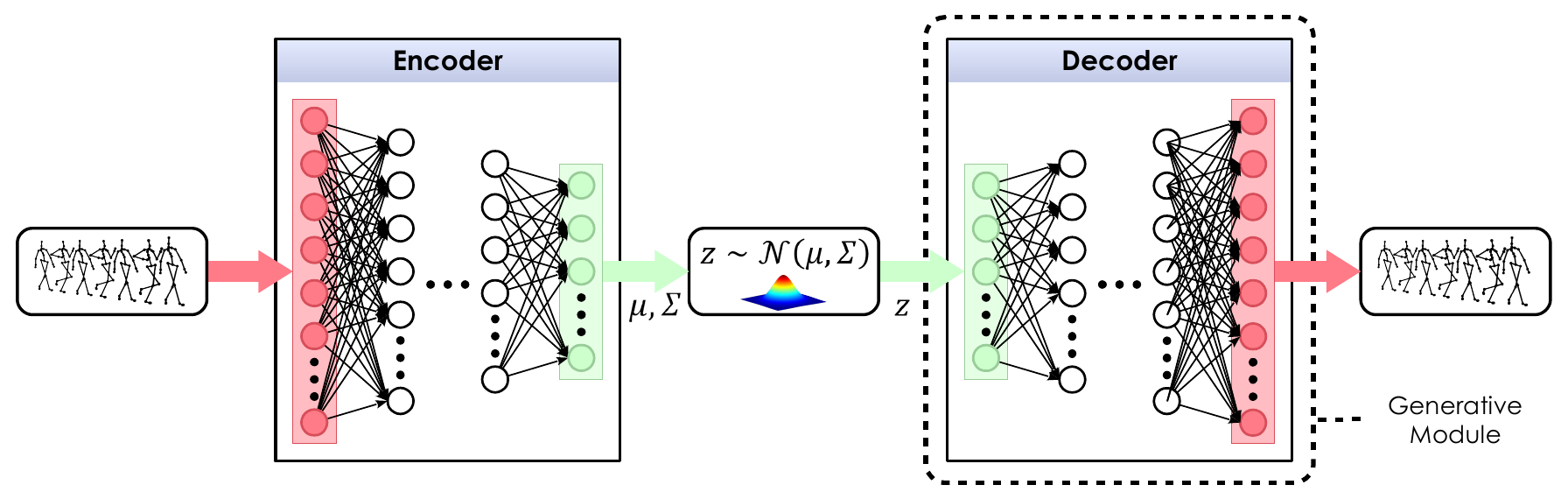}
    \caption{
        In a \acrFull{VAE} a motion sequence is mapped by an encoder to a random latent code that is constrained to follow a prior Gaussian distribution. This code is mapped back by a decoder to the input motion data. Once the full network is trained, feeding the decoder with samples drawn from the prior generates stochastic motion sequences.
    }
    \label{fig:VAE}
\end{figure}
\subsubsection{Variational Autoencoders}
    Structured as an encoder followed by a decoder, an autoencoder is a \acrShort{DNN} whose input and output represent the same data. The encoder output provides an intermediate latent code with a dimension often lower than the input data. Thus, autoencoders provide a scheme for non-linear dimensionality reduction. As illustrated in Figure~\ref{fig:VAE}, the distribution of the latent codes in a \acrFull{VAE}~\cite{ref:KW14} is further constrained to follow a predefined prior distribution, typically a multivariate normal distribution, endowing the autoencoder with a generative capability. Thus, \acrShortPlr{VAE} provide a convenient way to embed a stochastic component into the generation of human motion.

    Various approaches have been proposed to extend the \acrShort{VAE} framework to the modelling of temporal sequences. Toyer~\etal~\cite{ref:TC17} relied on a Deep Markov Model, essentially a \acrShort{VAE} in which the latent code is conditioned on its value at the previous time step. Habibie~\etal~\cite{ref:HH17} combined a \acrShort{VAE} and an \acrShort{RNN}. Motion generation is driven by the random latent code samples, as well as by control variables that constrain the trajectory and velocity of the character. The \acrShort{RNN} is conditioned on encodings on these control variables. At inference time, its cell state is initialised with the latent code value. The concatenation of cell state outputs at each time step is fed to the decoder to produce the synthesised motion sequence.

    Du~\etal~\cite{ref:DH19} built on the motion graph framework proposed by Min and Chai~\cite{ref:MC12}, in which motion sequences are represented as a graph of motion primitives. The segmentation of motion sequences into primitives is dependent on the type of motion and typically hand-crafted. Autoencoders learn embeddings for each primitive, and the latent codes for these embeddings are further encoded by Conditional \acrShortPlr{VAE} trained on dataset samples for the considered primitives. The conditioning of the primitive-specific \acrShortPlr{VAE} ensures that they reproduce the style of the input motion, which is encoded as a Gram matrix in the embedding space, following prior work in motion style transfer~\cite{ref:HS16} (see Section~\ref{subsection:StyleTransfer}). To synthesise a motion sequence, a path is determined in the motion graph based on user-defined trajectory controls, and motion primitives are generated along this path using the \acrShortPlr{VAE}.

    Yan~\etal~\cite{ref:YR18} and Aliakbarian~\etal~\cite{ref:AS20} relied on similar network architectures for stochastic motion prediction. Pose sequences are mapped to lower-dimensional features by an encoder and transformed back to motion data by a decoder. The \acrShort{VAE} operates on the encoded features, its output is fed to the decoder to synthesise the motion clips. Yan~\etal~\cite{ref:YR18} processed small pose sequences called \emph{motion modes} that capture short-term motion features. During training, their \acrShort{VAE} maps a pair of (past, future) mode features to a random latent code (encoder) then to a prediction of the future mode feature (decoder). It thereby captures the transition between the two modes. At inference time, the \acrShort{VAE} and \acrShort{RNN} decoders generate a stochastic prediction of the future mode, given a past conditioning mode and a draw of the \acrShort{VAE} random latent code. Aliakbarian~\etal~\cite{ref:AS20} argued that stacking the past sequence information and the random generating seed in a vector and feeding it to the decoding network leaves the possibility that the stochastic component is assigned low weights during training and is thus effectively ignored by the network. This concern is confirmed by experimental evidence. To avoid this, they proposed a \emph{mix-and-match perturbation} strategy and formed a vector by replacing randomly selected components of the past sequence feature by corresponding components of the \acrShort{VAE} latent code. Feeding this vector to the decoder forces it to account for both the past context and the stochastic input.

    \clearpage

\subsubsection{Generative Adversarial Networks}
    \begin{figure}[t]
    \centering
    \includegraphics[width=\linewidth]{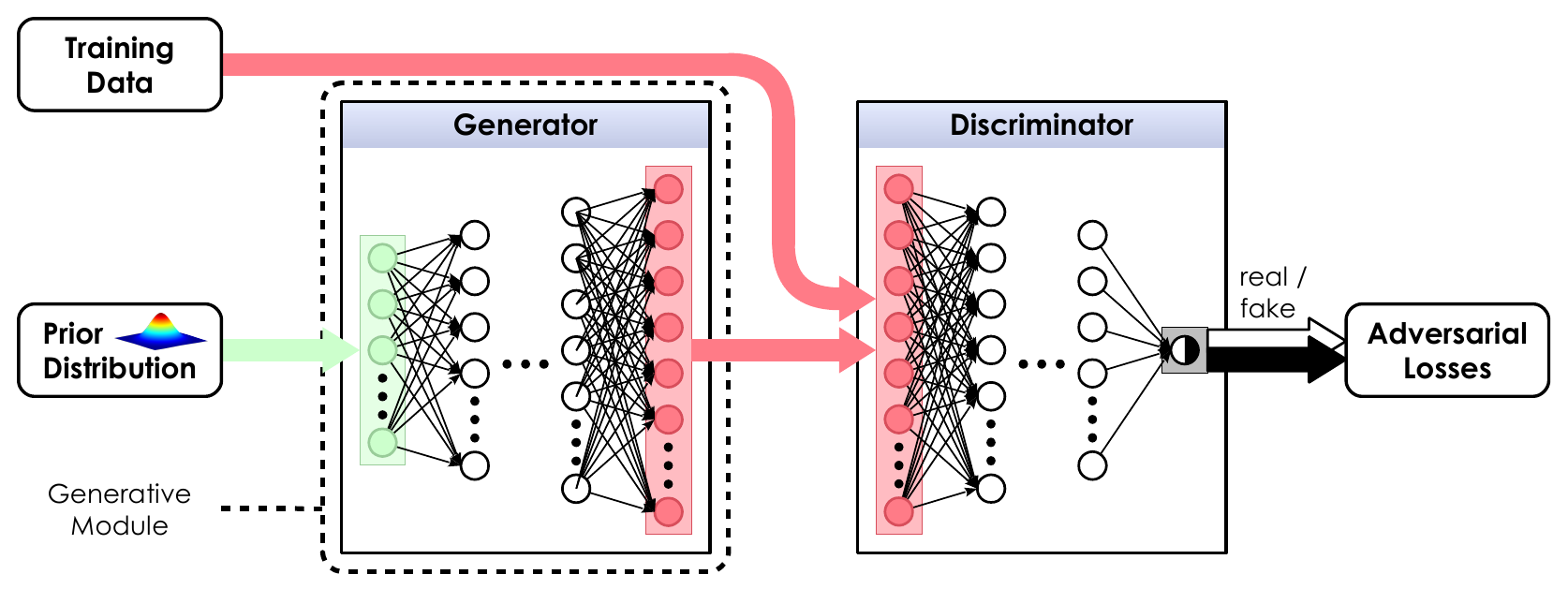}
    \caption{
        A \acrFull{GAN} consists of two networks: the generator produces ``fake'' motion samples from random seeds, the discriminator tries to differentiate them from ``real'' ones drawn from the training dataset. The two modules are trained jointly, aiming at an equilibrium where the generator outputs cannot be distinguished from the training data.
    }
    \label{fig:GAN}
\end{figure}
    \acrFullPlr{GAN} are a popular alternative to \acrShortPlr{VAE} for generating samples from random seeds. In a \acrShort{GAN}, new samples are synthesised by a generator network that operates in conjunction with a discriminator network (see Figure~\ref{fig:GAN}). The generator transforms a random seed drawn from a known prior distribution to a sample which aims to be similar to the contents of the training dataset, the discriminator assesses this similarity. The two networks are trained jointly with adversarial losses, the generator being driven to produce samples that the discriminator ultimately should not be able to distinguish from the training dataset samples.

    Barsoum~\etal~\cite{ref:BK18} pointed out that relying on a \acrShort{GAN} for long-term motion prediction has several intrinsic advantages. \acrShortPlr{GAN} do not suffer from the temporal error accumulation issues of \acrShortPlr{RNN} and will produce one instance of all possible outputs instead of averaging these possible outputs as \acrShort{ERD} networks tend to do. \acrShortPlr{GAN} also natively embed a stochastic component. Any motion generation network can be wrapped into a \acrShort{GAN} by adding a discriminator network to improve the plausibility of the generated sequences~\cite{ref:BK18, ref:WC21, ref:HY20}. Other works focused their contribution on optimising the architecture of the \acrShort{GAN} generator network to facilitate and improve its training. Wang~\etal~\cite{ref:WA20} proposed an adversarial autoencoder architecture~\cite{ref:MS15} where a \acrShort{GAN} enforces a prior on the distribution of the latent code of an autoencoder. Yan~\etal~\cite{ref:YL19} fed the generator with a sequence of random samples drawn from a Gaussian process with a predetermined covariance function. Through a series of purely convolutional modules made up of a spatio-temporal upsampling layer followed by a graph convolution on the skeleton features, they gradually increased the spatial and temporal resolution of their output to produce a pose sequence. Wang~\etal~\cite{ref:WY20} split their generator into separate spatial and temporal sub-networks: the lower layer maps the input random seed and conditioning action label via an \acrShort{RNN} to a sequence of low-dimensional latent codes that model temporal transitions. The upper layer decodes each latent code into a skeletal pose. The generator is regularised by a helper action classifier network through a cycle consistency constraint: the conditioning action label fed to the generator should agree with the result of the classification of the motion sequence it produces.

\subsubsection{Normalising Flows}
    Generative models based on \emph{Normalising Flows}~\cite{ref:DK14} synthesise samples by applying a composition of invertible elementary transforms to a latent variable drawn from a known prior distribution. Unlike in \acrShortPlr{GAN} or \acrShortPlr{VAE}, owing to the form of the elementary transforms, the probability density function of the generated samples can be computed in closed form. Thus, the generative network can be trained by maximum likelihood optimisation. Henter~\etal~\cite{ref:HA20} adapted this framework to the generation of motion sequences. Each transform layer in the generator network is conditioned by a control signal that encodes the past trajectory of the root joint and holds an \acrShort{LSTM} unit whose hidden state captures temporal dependencies.
    
\subsubsection{Miscellaneous}
    \paragraph*{Deterministic Generation.}
    Holden~\etal~\cite{ref:HS16} conditioned their motion generation approach on purely deterministic constraints that specify the trajectory and velocity of the character. They leveraged an autoencoder operating on temporal chunks of poses to learn a latent manifold of human motion. A separate feedforward network maps the autoencoder embedding to high-level, semantically interpretable controls of the motion. Generating high-dimensional motion embedding samples from a set of low-dimensional control parameters is severely under-constrained. To disambiguate the generation to the largest possible extent, the scope of the approach is restricted to human locomotion, and a comprehensive set of trajectory and foot contact constraints is imposed.

    \paragraph*{Deep Generative Models Sampling.}
    After training a deep generative model driven by an input random seed, motion sequences are often obtained by drawing independent samples from the seed and feeding them to the generator network. Yuan et Kitani~\cite{ref:YK20} proposed a better sampling strategy to enforce diversity in the generated sequences and cover minor modes of their distribution. To this end, they computed correlated random seeds by applying a set of affine transforms to a random variable drawn from a Gaussian distribution. The parameters of the affine transforms are learnt by a front-end network that can be conditioned on the same conditioning data used by the generative model. Each seed generates one motion sequence. Diversity among these sequences is enforced through a repulsion loss term that maximises the dissimilarity between the generated sequences. Another loss term enforces consistency of the random seeds with the prior distribution of the generative model. The relative weights of these two terms define a trade-off between the conflicting goals of diversity and fidelity to the training data distribution.
    
    % trick to avoid starting next section before table 2
    \vspace{2cm}

    \begin{table*}
    \setlength{\tabcolsep}{5.0pt}
    \scriptsize
    \centering
    \caption{
        Summary of the methods presented in Section \ref{section:MotionSynthesis}. Miscellaneous data includes hand-crafted, synthetic, proprietary, unspecified or other public datasets.
    }
    \begin{NiceTabular}{| c | c | *{3}{c} | c | *{7}{c} | *{10}{c} | c |}
        \CodeBefore
            \rowcolors{4}{tablerowcolor1}{tablerowcolor2} % Set alternating row colors
        \Body
        \hline
        & & && & & &&&&&& & &&&&&&&&& & \\
        
        % Headers
        \header{Reference}                              &
        \header{Code}                                   &
        \Block{1-3}{\header{Section}} &&                &
        \header{Framework}                              &
        \Block{1-7}{\header{Dataset}} &&&&&&            &
        \Block{1-10}{\header{Architecture}} &&&&&&&&&   &
        \header{Representation}                         \\
        
        & & && & & &&&&&& & &&&&&&&&& & \\
        
        % Sub-headers
        & % no sub-header (Reference)
        & % no sub-header (Code)
        \rotatebox{90}{\subheader{\ref{subsection:ShortTermPrediction} Short-term Prediction}}      &
        \rotatebox{90}{\subheader{\ref{subsection:LongTermPrediction} Long-term Prediction}}        &
        \rotatebox{90}{\subheader{\ref{subsection:DeepGenerativeModeling} Generative Synthesis}}    &
        & % no sub-header (Framework)
        \rotatebox{90}{\subheader{3DPW} \cite{ref:VH18}}                                            &
        \rotatebox{90}{\subheader{CMU \cite{ref:CM03}}}                                             &
        \rotatebox{90}{\subheader{Human3.6M \cite{ref:IP14}}}                                       &
        \rotatebox{90}{\subheader{HDM05} \cite{ref:MR07}}                                           &
        \rotatebox{90}{\subheader{Holden \etal \cite{ref:HS16}}}                                    &
        \rotatebox{90}{\subheader{NTU RGB+D\cite{ref:SL16}}}                                        &
        \rotatebox{90}{\subheader{Miscellaneous}}                                                   &
        \rotatebox{90}{\subheader{Fully Connected}}                                                 &
        \rotatebox{90}{\subheader{Convolutional}}                                                   &
        \rotatebox{90}{\subheader{Graph Convolutional}}                                             &
        \rotatebox{90}{\subheader{Recurrent}}                                                       &
        \rotatebox{90}{\subheader{Transformer}}                                                     &
        \rotatebox{90}{\subheader{Autoencoder}}                                                     &
        \rotatebox{90}{\subheader{RBM}}                                                             &
        \rotatebox{90}{\subheader{Variational Autoencoder}}                                         &
        \rotatebox{90}{\subheader{Adversarial}}                                                     &
        \rotatebox{90}{\subheader{Normalising Flows}}                                               &
        \\ % no sub-header (Representation)
    
        \hline
        \cite{ref:FL15}    & \useurl{code:FL15}    & × &   &   & DL        &   &   & × &   &   &   &   &   &   &   & × &   &   &   &   &   &   &          $\so{3}$           \\
        \cite{ref:JZ16}    & \useurl{code:JZ16}    & × &   &   & DL        &   &   & × &   &   &   &   &   &   &   & × &   &   &   &   &   &   &          $\so{3}$           \\
        \cite{ref:BB17}    & \useurl{code:BB17}    & × &   &   & DL        &   & × & × &   &   &   &   & × &   &   &   &   &   &   &   &   &   &          $\so{3}$           \\
        \cite{ref:MB17}    & \useurl{code:MB17}    & × &   &   & DL        &   &   & × &   &   &   &   &   &   &   & × &   &   &   &   &   &   &          $\so{3}$           \\
        \cite{ref:GW18a}   & \useurl{code:GW18a}   & × &   &   & DL        &   &   & × &   &   &   &   &   &   &   & × &   &   &   &   & × &   &          $\so{3}$           \\
        \cite{ref:GW18b}   & \useurl{code:GW18b}   & × &   &   & DL        &   &   & × &   &   &   &   &   &   &   & × &   &   &   &   &   &   &          Unknown            \\
        \cite{ref:LZ18}    & \useurl{code:LZ18}    & × &   &   & DL        &   & × & × &   &   &   &   &   & × &   &   &   &   &   &   & × &   &          $\so{3}$           \\
        \cite{ref:AK19}    & \useurl{code:AK19}    & × &   &   & DL        &   &   & × &   &   &   &   &   &   &   & × &   &   &   &   &   &   &          $\SO{3}$           \\
        \cite{ref:CA19}    & \useurl{code:CA19}    & × &   &   & DL        &   &   & × &   &   &   & × &   &   &   & × &   &   &   &   &   &   &         3D velocities       \\
        \cite{ref:GC19}    & \useurl{code:GC19}    & × &   &   & DL        &   & × & × &   &   &   &   &   &   &   & × &   &   &   &   &   &   &          $\so{3}$           \\
        \cite{ref:LiW19}   & \useurl{code:LiW19}   & × &   &   & DL        &   & × & × &   &   &   &   &   & × &   &   &   &   &   &   &   &   &          $\so{3}$           \\
        \cite{ref:ML19}    & \useurl{code:ML19}    & × &   &   & DL        & × & × & × &   &   &   &   &   &   & × &   &   &   &   &   &   &   &             DCT             \\
        \cite{ref:WA19}    & \useurl{code:WA19}    & × &   &   & DRL       &   &   & × &   &   &   &   &   &   &   & × &   &   &   &   & × &   &          $\so{3}$           \\
        \cite{ref:AA20}    & \useurl{code:AA20}    & × &   &   & DL        &   &   &   &   &   & × & × &   & × &   & × &   &   &   &   &   &   &         3D positions        \\
        \cite{ref:CH20}    & \useurl{code:CH20}    & × &   &   & DL        &   & × & × &   &   &   &   &   &   &   &   & × &   &   &   &   &   &             DCT             \\
        \cite{ref:CP20}    & \useurl{code:CP20}    & × &   &   & DL        &   &   &   &   &   &   & × &   &   &   & × &   &   &   &   &   &   &         3D positions        \\
        \cite{ref:CS20}    & \useurl{code:CS20}    & × &   &   & DL        & × & × & × &   &   &   &   &   & × & × &   &   &   &   &   & × &   &          $\so{3}$           \\
        \cite{ref:LC20}    & \useurl{code:LC20}    & × &   &   & DL        &   & × & × &   &   &   &   &   & × & × & × &   &   &   &   &   &   &          $\so{3}$           \\
        \cite{ref:LK20}    & \useurl{code:LK20}    & × &   &   & DL        &   & × & × &   &   &   &   &   & × & × &   &   &   &   &   &   &   &         3D positions        \\
        \cite{ref:ML20}    & \useurl{code:ML20}    & × &   &   & DL        & × &   & × &   &   &   & × &   &   & × &   &   &   &   &   &   &   &             DCT             \\
        \cite{ref:PF20}    & \useurl{code:PF20}    & × &   &   & DL        &   &   & × &   & × &   &   &   &   &   & × &   &   &   &   &   &   &         Quaternions         \\
        \cite{ref:ZP20}    & \useurl{code:ZP20}    & × &   &   & DL        &   & × & × &   &   &   &   &   & × &   &   &   &   &   &   &   &   &          Unknown            \\
        \cite{ref:BG21}    & \useurl{code:BG21}    & × &   &   & DL        &   & × & × &   &   &   &   &   &   & × &   &   &   &   & × &   &   &             DCT             \\
        \cite{ref:CS21}    & \useurl{code:CS21}    & × &   &   & DL        & × & × & × &   &   &   &   &   & × &   &   &   &   &   &   & × &   &          $\so{3}$           \\
        \cite{ref:LC21}    & \useurl{code:LC21}    & × &   &   & DL        &   & × & × &   &   & × &   &   & × & × & × &   &   &   &   &   &   &          $\so{3}$           \\
        \cite{ref:YK21}    & \useurl{code:YK21}    & × &   &   & DL        &   & × &   &   &   &   & × &   &   &   & × &   &   &   &   &   &   &           Ad hoc            \\
        \cite{ref:PG18}    & \useurl{code:PG18}    & × & × &   & DL        &   &   & × &   & × &   &   &   &   &   & × &   &   &   &   &   &   &         Quaternions         \\
        \cite{ref:XL19}    & \useurl{code:XL19}    & × & × &   & DL        &   &   & × &   &   &   &   &   &   &   & × &   &   &   &   &   &   &          Unknown            \\
        \cite{ref:GS17}    & \useurl{code:GS17}    &   & × &   & DL        &   &   & × &   & × &   &   &   &   &   & × &   &   &   &   &   &   &          Unknown            \\
        \cite{ref:TM18}    & \useurl{code:TM18}    &   & × &   & DL        &   &   & × &   &   &   &   &   &   &   & × &   &   &   &   &   &   &          $\so{3}$           \\
        \cite{ref:ZL18}    & \useurl{code:ZL18}    &   & × &   & DL        &   & × &   &   &   &   &   &   &   &   & × &   &   &   &   &   &   &         3D positions        \\
        \cite{ref:GM19}    & \useurl{code:GM19}    &   & × &   & DL        &   &   & × &   &   &   &   &   &   &   & × &   &   &   &   &   &   &          $\so{3}$           \\
        \cite{ref:HG19}    & \useurl{code:HG19}    &   & × &   & DL        &   &   & × &   &   &   &   &   & × &   &   &   &   &   &   & × &   &         3D positions        \\
        \cite{ref:KG19}    & \useurl{code:KG19}    &   & × &   & DL        &   & × & × &   &   &   &   &   &   &   & × &   & × &   &   & × &   &          $\so{3}$           \\
        \cite{ref:LiuW19}  & \useurl{code:LiuW19}  &   & × &   & DL        &   &   & × &   &   &   & × &   &   &   & × &   &   &   &   &   &   &          $\se{3}$           \\
        \cite{ref:WH19}    & \useurl{code:WH19}    &   & × &   & DL        &   & × &   & × & × &   & × &   &   &   & × &   &   &   &   &   &   &         3D positions        \\
        \cite{ref:YK19}    & \useurl{code:YK19}    &   & × &   & DL        &   &   &   &   &   &   & × & × &   &   &   &   &   &   &   &   &   &          Unknown            \\
        \cite{ref:CG20}    & \useurl{code:CG20}    &   & × &   & DL        &   &   &   &   &   &   & × &   & × &   &   & × &   &   & × &   &   &         3D positions        \\
        \cite{ref:GW20}    & \useurl{code:GW20}    &   & × &   & DL        &   &   &   &   &   &   & × &   &   &   & × &   & × &   & × &   &   &         Quaternions         \\
        \cite{ref:KA20}    & \useurl{code:KA20}    &   & × &   & DL        &   &   &   &   & × &   &   &   & × &   &   &   & × &   &   &   &   &         3D positions        \\
        \cite{ref:XX20}    & \useurl{code:XX20}    &   & × &   & DL        &   & × &   &   &   &   &   &   &   &   & × &   &   &   &   & × &   &        Euler Angles         \\
        \cite{ref:HH17}    & \useurl{code:HH17}    &   & × & × & DL        &   &   &   &   &   &   & × &   & × &   & × &   &   &   & × &   &   &         3D positions        \\
        \cite{ref:HY20}    & \useurl{code:HY20}    &   & × & × & DL        &   &   & × &   &   &   & × &   &   &   & × &   &   &   &   & × &   &         Quaternions         \\
        \cite{ref:WC21}    & \useurl{code:WC21}    &   & × & × & DL        &   & × &   &   &   &   & × &   &   &   & × &   &   &   &   & × &   &          Unknown            \\
        \cite{ref:TH06}    & \useurl{code:TH06}    &   &   & × & DL        &   & × &   &   &   &   & × &   &   &   &   &   &   & × &   &   &   &          $\so{3}$           \\
        \cite{ref:TH09}    & \useurl{code:TH09}    &   &   & × & DL        &   & × &   &   &   &   & × &   &   &   &   &   &   & × &   &   &   &          $\so{3}$           \\
        \cite{ref:CM11}    & \useurl{code:CM11}    &   &   & × & DL        &   & × &   &   &   &   &   &   &   &   &   &   &   & × &   &   &   &          $\so{3}$           \\
        \cite{ref:AL15}    & \useurl{code:AL15}    &   &   & × & DL        &   &   &   &   &   &   & × &   &   &   &   &   &   & × &   &   &   &          $\so{3}$           \\
        \cite{ref:HS16}    & \useurl{code:HS16}    &   &   & × & DL        &   &   &   &   & × &   &   &   & × &   &   &   & × &   &   &   &   &         3D positions        \\
        \cite{ref:TC17}    & \useurl{code:TC17}    &   &   & × & DL        &   &   &   &   &   & × & × &   &   &   & × &   &   &   &   &   &   &         2D positions        \\
        \cite{ref:BK18}    & \useurl{code:BK18}    &   &   & × & DL        &   &   & × &   &   & × &   &   &   &   & × &   &   &   &   & × &   &         3D positions        \\
        \cite{ref:YR18}    & \useurl{code:YR18}    &   &   & × & DL        &   &   & × &   &   &   & × &   &   &   & × &   &   &   & × &   &   &         3D positions        \\
        \cite{ref:DH19}    & \useurl{code:DH19}    &   &   & × & DL        &   & × &   & × &   &   & × &   & × &   &   &   &   &   & × &   &   &         3D positions        \\
        \cite{ref:YL19}    & \useurl{code:YL19}    &   &   & × & DL        &   &   &   &   &   & × & × &   &   & × &   &   &   &   &   & × &   &          Unknown            \\
        \cite{ref:AS20}    & \useurl{code:AS20}    &   &   & × & DL        &   & × & × &   &   &   &   &   &   &   & × &   &   &   & × &   &   &         Quaternions         \\
        \cite{ref:HA20}    & \useurl{code:HA20}    &   &   & × & DL        &   & × &   & × & × &   & × &   &   &   & × &   &   &   &   &   & × &         3D positions        \\
        \cite{ref:WA20}    & \useurl{code:WA20}    &   &   & × & DL        &   &   &   &   &   &   & × &   &   &   & × &   & × &   &   & × &   &        Euler Angles         \\
        \cite{ref:WY20}    & \useurl{code:WY20}    &   &   & × & DL        &   &   & × &   &   & × &   &   &   &   & × &   &   &   &   & × &   &          Unknown            \\
        \cite{ref:YK20}    & \useurl{code:YK20}    &   &   & × & DL        &   &   & × &   &   &   & × &   &   &   & × &   &   &   & × &   &   &         3D positions        \\
        \hline
    \end{NiceTabular}
    \label{tab:synthesis}
\end{table*}
    \begin{figure*}
	\centering
	\mbox{} \hfill
	\includegraphics[width=0.74\linewidth]{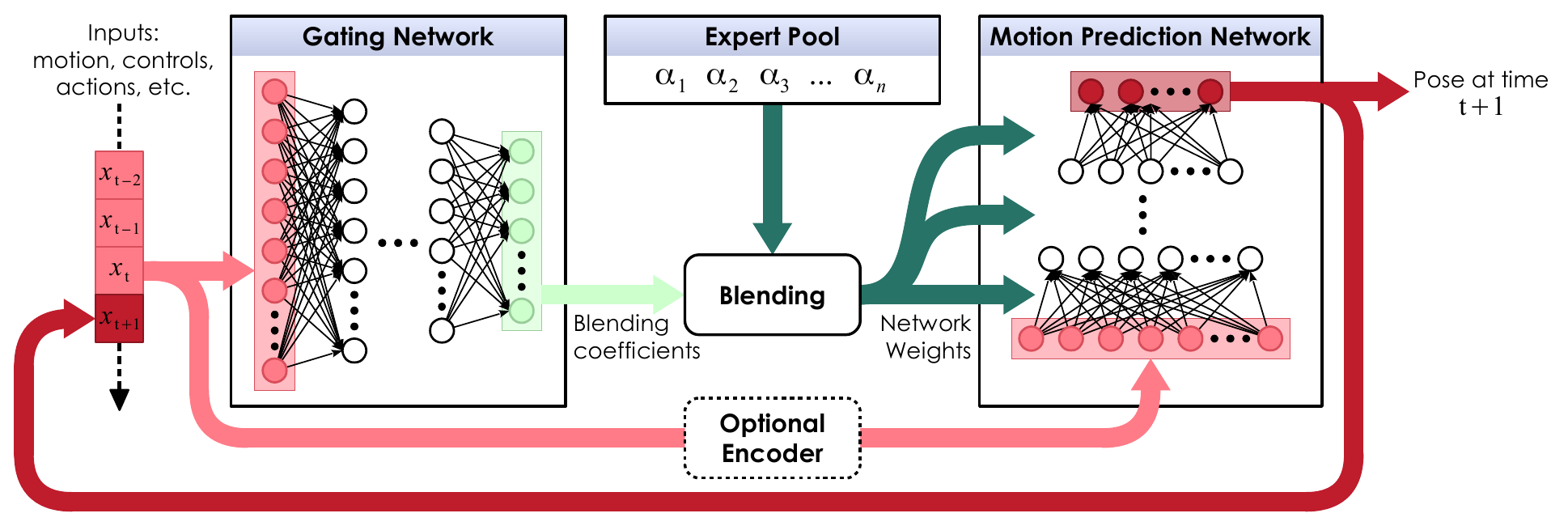}
	\hfill \mbox{}
    \caption{
        Overall idea of the mixture-of-experts scheme used in~\cite{ref:ZS18,ref:SZ19,ref:SZ20,ref:LZ20}: a gating network first computes blending weights for a number of expert networks, based on a number of features extracted from the current frame $x_t$. Each expert specialises in a particular movement. They are then blended together to dynamically compute the weights of a motion prediction network, which outputs information relevant to the next frame $x_{t+1}$ including the pose of the animated character. This output is typically fed back into the network (autoregression) at time $t+1$.
    }
    \label{fig:MoE}
\end{figure*}
    \section{Character Control} \label{section:CharCtrl}
    Controlling character motions that react naturally to user inputs, while accounting for environment constraints as well as biomechanical limitations, is another challenge involved in creating believable virtual characters. In this section, we explore this topic through three main axes: how to make characters move and interact with the virtual environment using kinematic (Section~\ref{subsection:KinCharCtrl}), physical (Section~\ref{subsection:PhysCharCtrl}) or biomechanical (Section~\ref{subsection:BioCharCtrl}) control. Finally, we summarise the methods presented in this section in Table~\ref{tab:control}.

\subsection{Kinematics-based} \label{subsection:KinCharCtrl}
    Kinematic approaches typically produce motions as joint angles, based on a set of motion examples and high-level controls (\eg user inputs, interactions with the environment). Because of the requirement of generating motions in an online fashion when controlling characters in video games or other interactive applications, \acrShortPlr{RNN} and other autoregressive models are often considered to be more appropriate than \acrShortPlr{CNN}, as the future pose is predicted from the previous motion as well as a control signal. For instance, Lee~\etal~\cite{ref:LL18} used a four-layer \acrShort{LSTM} model for controlling characters playing basketball and tennis. However, as mentioned in Section~\ref{subsection:LongTermPrediction}, such approaches often tend to fail in the long run, as errors in the prediction are fed back into the input and accumulate, eventually either converging to an average pose or introducing high frequency artifacts. According to Starke~\etal~\cite{ref:SZ19}, these models also often suffer from low responsiveness due to the large variation of the memory state in the case of interactive character control, as the internal memory state is high dimensional.

    % Control phase of the motion
    To overcome these limitations, Holden~\etal~\cite{ref:HK17} proposed the use of a specialised architecture called \acrFull{PFNN}, which provides the phase variable to represent the progression of the motion. In their seminal work, the phase is defined based on alternating foot contacts, and used to generate the weights of the regression network at each frame. A trade-off between compactness and runtime speed can then be achieved by precomputing the phase-function for a number of fixed intervals, then interpolating the precomputed elements at runtime. One major limitation of \acrShort{PFNN} is that phase functions need to be manually defined, which can be in some cases extremely complex~\cite{ref:ZS18,ref:SZ20}. Zhang~\etal~\cite{ref:ZS18} therefore proposed to rely on a mixture-of-experts scheme to dynamically compute the weights of a motion prediction network (see Figure~\ref{fig:MoE} for an illustration of the general concept). In their architecture, a gating network first computes blending weights for a number of expert networks, each specialising in a particular movement. This approach was first demonstrated for creating complex quadruped character controllers and then extended by Starke~\etal~\cite{ref:SZ19} to compute goal-directed series of motions and transitions, while potentially interacting with the environment. The idea was pushed one step further through the use of local motion phases~\cite{ref:SZ20}, which are defined based on how each body part contacts external objects. Unlike previous approaches where different actions are considered to be synchronised by a single global phase variable, their approach describes each motion by a set of multiple independent and local phases for each bone. It then enables neural networks to learn asynchronous movements of each bone, as well as its interaction with external elements of the virtual environment. While the previous approaches relied on autoregressive \acrShortPlr{DNN} to generate controlled character motions, Ling~\etal~\cite{ref:LingZ20} demonstrated that a \acrShort{VAE} using a similar mixture-of-experts scheme is also viable to produce stable high-quality human motions, while being usable in a \acrShort{DRL} context to produce goal-directed motions.
    
    % Generate expressive variations or style
    Simultaneously, a few approaches explored the creation of controllable expressive human motions from high-level semantic factors. For instance, Alemi and Pasquier~\cite{ref:AP17} trained a \acrShort{FCRBM} on a dataset of motion capture data containing movements from different subjects, expressions, and trajectories. This model can then be used to generate modulated walking movements in real time. Mason~\etal~\cite{ref:MS18} explored a similar question from the perspective of generating characters moving in different styles when there is little data available for a new style and proposed a few-shot learning approach. The goal of few-shot learning is to learn (part of) a model able to generalise out of a single or very few examples. In their work, Mason~\etal~\cite{ref:MS18} adapted a pre-trained \acrShort{PFNN} (modelling style-independent components of the motions), coupled with a set of residual adapters (modelling style-dependent components) learned separately for each new style.
 
    % Multiple character interactions
    While some of the approaches mentioned above have been demonstrated to be compatible with multi-character control, such character interactions are typically handled by directly including information about the other characters' relative positions~\cite{ref:SZ20}, or indirectly through information about objects both characters are interacting with and the action to be performed~\cite{ref:LL18,ref:SZ20}. Wang~\etal~\cite{ref:WC17} also proposed to generate character interactions based on the history motion data of both characters, relying on a variant of the \acrShort{ERD} architecture to improve animation stability, where multiple \acrShort{LSTM} layers constitute the recurrent network.
    
    % Leveraging DL with traditional animation approaches 
    Finally, despite the impressive advances made by the aforementioned methods, traditional animation approaches are still commonly used in animation pipelines to control human characters because of the quality of the motions produced. However, novel approaches, such as \emph{Learned Motion Matching}~\cite{ref:HK20}, have recently begun to be explored with the goal of breaking down and replacing individual components of animation algorithms by individual specialised neural networks. They balance the advantages of more traditional approaches with the scalability of neural network based models.

\subsection{Physics-based} \label{subsection:PhysCharCtrl}
    Physics-based approaches generate animations in agreement with physical laws. All but a few methods in this section rely on multibody simulations for physical coherence. \acrShort{DL} has also been combined with optimisation-based motion control to generate physically coherent animations. Early works such as Grzeszczuk~\etal~\cite{ref:GT98} or Peng~\etal~\cite{ref:PB16} pioneered the use of deep learning in physically realistic animation. While these were limited to animals, humans were quickly considered as well.

\begin{figure}
    \centering
    \includegraphics[width=\linewidth]{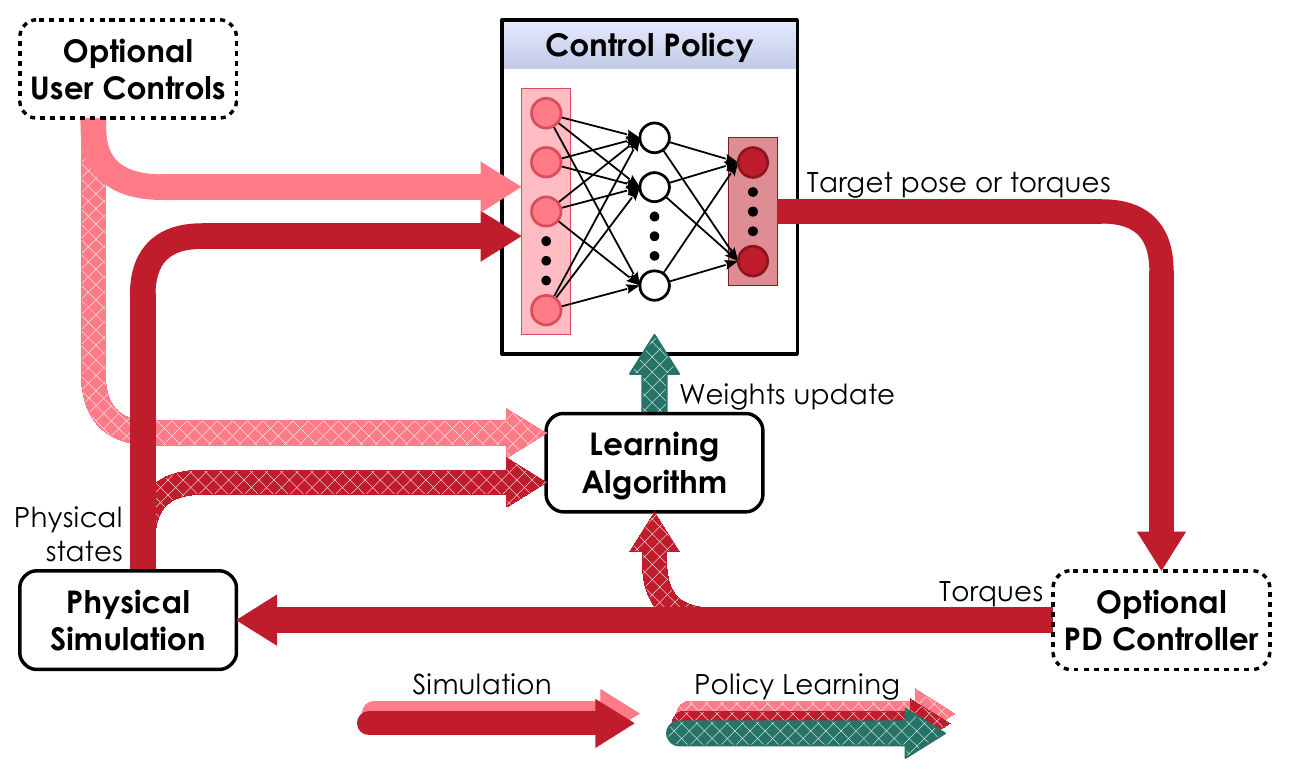}
    \caption{
        Illustration of the typical architecture for animation using multi-body simulations. A \acrFull{DNN} takes as input information about the state of the character, its surroundings and optionally user inputs such as the desired direction. It then generates either joint torques or target poses fed to a \acrFull{PD} controller. Due to the feedback loop, the model is trained by \acrFull{RL}.
    }
    \label{fig:control_loop}
\end{figure}
\paragraph*{Multibody Simulations.}
    The actuators of a multibody system are typically driven by a \acrShort{DNN} trained by \acrFull{RL}, where human characters commonly have between 22 and 62 \acrShortPlr{DOF}. In a nutshell, \acrFull{RL} is a framework to learn to control a system by maximising a reward signal. Therefore designing the reward is an important aspect of \acrShort{RL} algorithms. The learned model (\ie the agent) takes as inputs observations (\ie the states) from the system and outputs actions that impact the system. In the animation context, the states are often a combination of proprioceptive information (\eg positions, angles, velocities and angular velocities of the joints and the root of the body, end-effector contacts), the phase of the gait, as well as external information (\eg terrain height, interacting object information) and task information (\eg path to follow, direction of the goal). Similarly, the reward is often a combination of physics-based penalties (\eg losing balance, using more energy to produce a motion) and task-specific rewards (\eg matching a reference motion, satisfying a desired velocity for locomotion). Typically the actions are used to control the character and consist of either torques applied at the joints of the multibody system or target angles for said joints that are converted to torques by \acrFull{PD} controllers. The dynamics of the multibody system is then computed by a physical simulator (\eg \emph{Bullet} or \emph{MuJoCo}), as illustrated in Figure~\ref{fig:control_loop}.

    Training a \acrShort{DRL} model is different from supervised \acrShort{DL}. Rather than using a labeled dataset, the model is typically trained by interacting with the environment and improves at the same time. The most popular \acrShort{DRL} algorithm for animation is \emph{Proximal Policy Optimisation} (PPO)~\cite{ref:SW17}. Common practices to improve training include imitating reference motion capture data; using two controllers: a low-level one for locomotion or low-level actions and a higher one for long term goals; enforcing symmetry of the motion; using a curriculum, that is, learning increasingly difficult tasks; clever state initialisation; early termination based on the lack of end-effector contacts or low rewards received.

\paragraph*{Optimisation-based Motion Control.}
    These methods first compute a set of constraints (\eg footstep positions) based on the environment and a simplified model of the character dynamics (\eg an inverted pendulum model). The animation is then generated by solving a constrained optimisation problem combining these trajectory and character-level constraints. \acrShort{DL} is used here to accelerate and approximate one or several computationally demanding steps.

\subsubsection{Character Locomotion}
    % 2D
    Locomotion was the first task addressed by multibody simulations and \acrShort{DRL}. Peng and van de Panne~\cite{ref:PV17} first compared different action spaces for planar locomotion of bipeds and animals: torques, target joint angles, target joint angle velocities and muscle-activations. They observed that  using target joint angles as action space performs well and is the most robust in all cases. It is also the fastest to train in 5 out of 7 cases. Biped locomotion was then quickly considered in \DDD. Several works have developed two-level controllers. For instance, Peng~\etal~\cite{ref:PB17} used a high-level controller to output future footsteps, while a low-level controller trained to imitate reference motions is used to actuate the character. Merel~\etal~\cite{ref:MT17} also constructed low-level controllers that imitate reference motion clips and control actuation. However, multiple low-level controllers are trained using generative adversarial imitation learning. Each controller is trained on a specific skill such as walking, turning, running, and the high-level controller selects the most appropriate instance. Similarly, Merel~\etal~\cite{ref:MA19} trained a character that only perceives its environment through egocentric vision, where a \acrShort{LSTM}-based vision-driven high-level controller learns to select among a number of low-level controllers that have learned to imitate short motion clips. Low-level policies can also be combined~\cite{ref:PC19} by multiplying their distribution over actions and exponentially weighting them by a weight computed by another neural network. Bergamin~\etal~\cite{ref:BC19} proposed to select short motion capture clips using a motion matching approach. Target angles are then extracted from the selected clip and a subset of these angles are corrected by a \acrShort{DRL} model to stabilise the resulting motion. Chentanez~\etal~\cite{ref:CM18} used a similar approach to reproduce a reference clip provided as input, \ie without including a mechanism for automatic clip selection.
    
    % 3D & curriculum
    Improving physical character locomotion has also been explored through the use of curricula, training scenarii where the agent is requested to perform a sequence of typically increasingly difficult tasks. For instance, building a curriculum of environments~\cite{ref:HT17} can help to train an agent that is able to move forward and run, jump, crouch and turn as needed. Xie~\etal~\cite{ref:XL20} also compared four environment generating curricula to learn to walk over stepping stones. The resulting policy can then be used to walk over difficult terrain where footstep locations are given. The state includes the height of the pelvis with respect to the lowest foot and the reward includes incentives for placing the foot at the center of a stone and progressing towards the next one. Curricula can also be constructed by assisting the character. Yu~\etal~\cite{ref:YT18} trained a model to generate symmetric and low-energy motion to create more realistic motion without using motion capture data. Symmetry is encouraged by an additional loss function. The curriculum is automatically created by including and reducing forces that help with forward motion and left-right balance, leading to different gaits emerging for different target velocities. Abdolhosseini~\etal~\cite{ref:AL19} also addressed learning symmetric gaits by either duplicating observed data by symmetry or by enforcing symmetry in the network. As the approaches discussed so far typically train one model for each morphology, Won and Lee~\cite{ref:WL19} developed a parametric \acrShort{DRL} controller allowing the shape of the character to be modified during locomotion. At the heart of the training algorithm is an adaptive body shape sampling mechanism that progressively restricts the agent to body shapes it has not mastered yet. 
    
    % misc
    Other types of approaches have also been explored in the community. Rajam\"aki and H\"am\"al\"ainen~\cite{ref:RH17, ref:RH19} developed a sampling-based approach for locomotion: a \emph{Monte Carlo Tree Search} explores possible actions to maximise rewards over multiple future time steps. In the tree, each child node is one time step ahead of its parent. The impact of a sequence of candidate actions is evaluated by physical simulations. The expansion of the tree (that is, the trajectories explored) is driven in part by \acrShort{GAN} models. After a set budget, the most promising action is selected for the next time step. Babadi~\etal~\cite{ref:BN19} later leveraged Rajam\"aki and H\"am\"al\"ainen's approach to train a \acrShort{DRL} network that directly predicts the next action, without looking ahead, as most models in this section do. The Monte Carlo Tree Search approach is used to quickly produce adequate reference motions. Plausible states are extracted from these motions and serve as initial states during training of a \acrShort{DRL} model. The authors highlighted a significant reduction in training time.
    
    % DL-based optimisation techniques
    A few approaches also explored the use of \acrShort{DL}-based optimisation techniques to physically animate walking characters. Kwon~\etal~\cite{ref:KL20} first computed rough successive center of mass and footstep positions using an inverted pendulum on a cart model, refined these estimates and added contact forces, timings and durations using a centroidal dynamics model, then modified these estimates to take external forces into account and finally generated a full-body motion using an \acrShort{IK} solver. A \acrShort{DNN} was trained to approximate these last two steps for real time performance. Mordatch~\etal~\cite{ref:ML15} trained an \acrShort{RNN} model to predict the next pose of a character to generate physically realistic trajectories. Their model was not trained to reproduce the output of a given dynamical system. Instead, they jointly generated physically realistic trajectories and trained the network on these trajectories, which were generated by solving an optimisation problem balancing physical realism (equations of motion, non-penetration, and force complementarity) and the stability of the neural network.

\subsubsection{Other Motions}
    More complex types of motion have also been considered in addition to or instead of locomotion. Peng~\etal~\cite{ref:PA18} trained \acrShort{DRL} models to imitate motion clips. Their approach is applied to locomotion and various acrobatics and martial arts skills. The reward of the training algorithm includes both an imitation term and a task specific one. The model is trained incrementally, starting by the end of the motion. Three approaches are proposed to sequence skills: 1) using the closest clip to compute the imitation reward, 2)~using a skill selection vector and 3) learning one policy per skill and using at every time step the policy expected to produce the highest rewards based on the current state. Peng~\etal~\cite{ref:PK18} developed a similar approach based on videos rather than motion clips. For each video clip, \DD and \DDD poses are extracted and a \DDD reference pose trajectory is reconstructed by optimising in the latent space of an autoencoder. Furthermore, the dataset is augmented by rotations and a curriculum is automatically constructed by another agent trained to propose initial states. Ma~\etal~\cite{ref:MY21} proposed to impose space-time bounds around samples of a trajectory rather than an imitation reward to learn from reference motions. These bounds only allow small deviations in joint angles and in the positions of the center of mass and end effectors. Learning is possible using only a binary reward tied to the respect of these bounds and early termination. Furthermore, to learn different styles, they proposed different additional rewards to steer the model towards animations resulting in a different energy level or convex hull volume. They also sampled initial states to favour states from which the current model achieves relatively lower rewards. Ranganath~\etal~\cite{ref:RX19} compressed the action space using principal or independent component analysis, reducing the size of the output of the \acrShort{DRL} model. Their method is otherwise similar to~\cite{ref:PA18}. 
    
    While some approaches learn a different model for each motion clip, Merel~\etal~\cite{ref:MH19} trained a single controller able to execute many skills. They trained a \acrShort{VAE} embedding current and future states from motion capture clips and decoding the action from the embedding and the current state. They then trained a high-level controller to set the value of the latent variable based on the task. On a side note, Wang~\etal~\cite{ref:WM17} learned a similar embedding (although based on previous rather than future states), but used it only for reproducing a given clip, not to automatically generate new animations. Won~\etal~\cite{ref:WG20} took as input a motion controller and trained a controller for a physically simulated character that can reproduce all motions of the input controller. At runtime, the trained controller is used to track such a motion. They used a two-level model by clustering generated motions, training one model by cluster and combining them in a mixture. They demonstrated their results on walking, running, acting and performing various dance styles. They also compared additive to multiplicative combination of different reward terms and chose the latter, arguing that it forces all reward terms to be high and that they obtained better results with complex motions such as dancing.

\subsubsection{Interactions}
    \acrShort{DRL} has also been used to animate characters interacting with other characters and complex objects. On a side note, some methods already described above have been applied to simple interactions such as pushing an object~\cite{ref:PB17, ref:PK18, ref:PC19}. Liu and Hodgins~\cite{ref:LH17} tackled animating a character walking on a ball, balancing on a bongo board or skateboarding by generating control fragments from motion capture data and training a controller  to select a generated control fragment at runtime. A control fragment is a short (0.1s) segment of motion capture data and an associated linear control policy. The same authors~\cite{ref:LH18} later addressed basketball dribbling. One controller is first learned for the body and legs. Then, a second controller is trained for the arms. Both are based on control fragments extracted from motion capture data. For these two controllers, the actions are corrected by a linear model and a \acrShort{DRL} model, respectively. Transitions between control fragments are also learned. Basketball has also been addressed by Park~\etal~\cite{ref:PR19}, together with obstacle racing, chicken hopping, fighting with other characters and various other skills. A \acrShort{DRL} controller is trained to correct target poses generated by a \acrShort{RNN} and then fed to \acrShort{PD} controllers. The \acrShort{RNN} uses multi-objective character control similar to~\cite{ref:LL18}. Pushing character interaction much further, Haworth~\etal~\cite{ref:HB20} used multi-agent \acrShort{DRL} to train a two-level model for crowd animation inspired by Peng~\etal~\cite{ref:PB17}, where the low-level controls one character based on two future footstep targets provided by the high-level controller. The latter is specific for each character whereas the former is shared. The input of the high-level controller includes a velocity field of nearby objects and agents.
    
    Merel~\etal~\cite{ref:MT20} extended their previous work~\cite{ref:MA19} to object manipulation tasks from egocentric vision. It involves a hierarchical controller based on a latent space (similar to~\cite{ref:MH19}), learning tasks step by step and task variations. Notably, the object is not part of the state. The character can move objects between shelves or catch and throw balls.
    
    Interactions with objects of the scene have also been explored with respect to clothes, in order to animate an upper human body dressing up~\cite{ref:CY18}. The key ideas include using haptic information in the state, dividing the task into subtasks, using the end states of one subtask as initial states for training the next one and including cloth deformation in the reward.
    
\subsection{Biomechanics-based} \label{subsection:BioCharCtrl}
    Concurrently to joint-actuated approaches for motion control, neural networks have also been explored in the context of musculoskeletal simulation. The general idea is that the complexity of traditional musculoskeletal models usually leads to costly simulations, and that neural network controllers can provide an efficient solution by learning biomimicking controls using data simulated offline with a detailed biomechanical model, to efficiently output optimal muscle activations at runtime. 
    
    Lee and Terzopoulos~\cite{ref:LT06} originally applied this idea to design novel neuromuscular control models of the human head-neck system (comprising of 7 cervical vertebrae and 72 neck muscles) learned in supervised frameworks, first using shallow neural networks generating both pose and tone control signals for the movements of the head. Then, Nakada and Terzopoulos~\cite{ref:NT15} improved head stability in a larger range of situations using deep stacked denoising autoencoders. Nakada~\etal~\cite{ref:NZ18} further generalised the concept to full-body human biomechanical simulation (193 bone model actuated by a total of 823 muscles) by designing a sensorimotor control system made of 20 fully connected \acrShortPlr{DNN}. In their framework, 5 \acrShortPlr{DNN} per retina are responsible for extracting visual information required to direct eye/head, arm and leg movements, while 10 additional \acrShortPlr{DNN} are responsible for the neuromuscular control of the 216 neck muscles, of the 29 muscles of each arm, and of the 39 muscles of each leg. More specifically, the limbs and head are driven by one voluntary motor \acrShort{DNN} each, responsible for generating efferent activation signals for the corresponding muscles to execute realistic controlled movements, and one reflex motor \acrShort{DNN} each, responsible for computing muscle activation adjustments due to low-level muscle dynamics. Lee~\etal~\cite{ref:LP19} also explored full-body musculoskeletal simulation using \acrShort{DRL}, with a two-level approach presented to learn simultaneously trajectory mimicking and muscle coordination.
    
    Finally, motion realism brought by such musculoskeletal approaches generally comes at the expense of complex modelling and costly simulations, which might explain to some extent why joint-actuation approaches are still most commonly used. To bridge the gap between both worlds, and \enquote{generate human-like motion comparable to muscle-actuation models, while retaining the benefit of simple modelling and fast computation offered by joint-actuation models}, Jiang~\etal~\cite{ref:JV19} proposed an approach to formulate the optimal control problem in the joint-actuation space while having an equivalent solution to the same problem in the muscle-actuation space. Both the metabolic energy function and state-dependent torque limits expressed in the joint-actuation space are approximated using fully connected \acrShortPlr{DNN} and learned in a \acrShort{DL} framework, and then used in conjunction with a policy learning algorithm built upon previous work~\cite{ref:YT18} (See section~\ref{subsection:PhysCharCtrl}).

\begin{table*}
    \setlength{\tabcolsep}{7.3pt}
    \scriptsize
    \centering
    \caption{
        Summary of the methods presented in Section \ref{section:CharCtrl}. Miscellaneous data includes hand-crafted, synthetic, proprietary, unspecified or other public datasets.
    }
    \begin{NiceTabular}{| c | c | *{3}{c} | c | *{3}{c} | *{5}{c} | *{6}{c} |}
        \CodeBefore
            \rowcolors{4}{tablerowcolor1}{tablerowcolor2} % Set alternating row colors
        \Body
        \hline
        & & && & & && & &&&& & &&&&& \\
        
        % Headers
        \header{Reference}                              &
        \header{Code}                                   &
        \Block{1-3}{\header{Section}} &&                &
        \header{Framework}                              &
        \Block{1-3}{\header{Dataset}} &&                &
        \Block{1-5}{\header{Architecture}} &&&&         &
        \Block{1-6}{\header{Model \& Simulation}} &&&&& \\
        
        & & && & & && & &&&& & &&&&& \\
        
        % Sub-headers
        & % no sub-header (Reference)
        & % no sub-header (Code)
        \rotatebox{90}{\subheader{\ref{subsection:KinCharCtrl} Kinematics-based}}   &
        \rotatebox{90}{\subheader{\ref{subsection:PhysCharCtrl} Physics-based}}     &
        \rotatebox{90}{\subheader{\ref{subsection:BioCharCtrl} Biomechanics-based}} &
        & % no sub-header (Framework)
        \rotatebox{90}{\subheader{CMU \cite{ref:CM03}}}                             &
        \rotatebox{90}{\subheader{Holden \etal \cite{ref:HS16}}}                    &
        \rotatebox{90}{\subheader{Miscellaneous}}                                   &
        \rotatebox{90}{\subheader{Fully Connected}}                                 &
        \rotatebox{90}{\subheader{Recurrent}}                                       &
        \rotatebox{90}{\subheader{RBM}}                                             &
        \rotatebox{90}{\subheader{Variational Autoencoder}}                         &
        \rotatebox{90}{\subheader{Adversarial}}                                     &
        \rotatebox{90}{\subheader{Simulator}}                                       &
        \rotatebox{90}{\subheader{Dimensionality}}                                  &
        \rotatebox{90}{\subheader{Humanoid(s)}}                                     &
        \rotatebox{90}{\subheader{Other biped(s)}}                                  &
        \rotatebox{90}{\subheader{Monoped(s)}}                                      &
        \rotatebox{90}{\subheader{Quadruped(s)}}                                    \\
        
        \hline
        \cite{ref:AP17}    & \useurl{code:AP17}    & × &   &   & DL        &   &   & × &   &   & × &   &   &         & 3D & × &   &   &   \\
        \cite{ref:HK17}    & \useurl{code:HK17}    & × &   &   & DL        &   &   & × & × &   &   &   &   &         & 3D & × &   &   &   \\
        \cite{ref:WC17}    & \useurl{code:WC17}    & × &   &   & DL        & × &   & × &   & × &   &   &   &         & 3D & × &   &   &   \\
        \cite{ref:LL18}    & \useurl{code:LL18}    & × &   &   & DL        &   &   & × &   & × &   &   &   &         & 3D & × &   &   &   \\
        \cite{ref:MS18}    & \useurl{code:MS18}    & × &   &   & DL        & × &   &   & × &   &   &   &   &         & 3D & × &   &   &   \\
        \cite{ref:ZS18}    & \useurl{code:ZS18}    & × &   &   & DL        &   &   & × & × &   &   &   &   &         & 3D &   &   &   & × \\
        \cite{ref:SZ19}    & \useurl{code:SZ19}    & × &   &   & DL        &   &   & × & × &   &   &   &   &         & 3D & × &   &   &   \\
        \cite{ref:HK20}    & \useurl{code:HK20}    & × &   &   & DL        &   &   & × & × &   &   &   &   &         & 3D & × &   &   &   \\
        \cite{ref:LingZ20} & \useurl{code:LingZ20} & × &   &   & DL \& DRL &   &   & × & × &   &   & × &   &         & 3D & × &   &   &   \\
        \cite{ref:SZ20}    & \useurl{code:SZ20}    & × &   &   & DL        &   &   & × & × &   &   &   & × &         & 3D & × &   &   &   \\
        \cite{ref:ML15}    & \useurl{code:ML15}    &   & × &   & DL        &   &   & × &   & × &   &   &   & MuJoCo  & 3D & × &   &   & × \\
        \cite{ref:HT17}    & \useurl{code:HT17}    &   & × &   & DRL       &   &   &   &   & × &   &   &   & MuJoCo  & 3D & × & × &   & × \\
        \cite{ref:LH17}    & \useurl{code:LH17}    &   & × &   & DRL       &   &   &   & × &   &   &   &   & ODE     & 3D & × &   &   &   \\
        \cite{ref:MT17}    & \useurl{code:MT17}    &   & × &   & DRL       & × &   &   & × &   &   &   & × & MuJoCo  & 3D & × &   &   &   \\
        \cite{ref:PB17}    & \useurl{code:PB17}    &   & × &   & DRL       &   &   & × & × &   &   &   &   & Bullet  & 3D &   & × &   &   \\
        \cite{ref:PV17}    & \useurl{code:PV17}    &   & × &   & DRL       &   &   & × & × &   &   &   &   & Unknown & 2D &   & × &   & × \\
        \cite{ref:RH17}    & \useurl{code:RH17}    &   & × &   & DRL       &   &   &   & × &   &   &   &   & ODE     & 3D & × &   & × & × \\
        \cite{ref:WM17}    & \useurl{code:WM17}    &   & × &   & DRL       & × &   &   &   & × &   & × & × & MuJoCo  & 3D & × & × &   &   \\
        \cite{ref:CM18}    & \useurl{code:CM18}    &   & × &   & DRL       &   & × &   & × &   &   &   &   & MuJoCo  & 3D & × &   &   &   \\
        \cite{ref:CY18}    & \useurl{code:CY18}    &   & × &   & DRL       &   &   &   & × &   &   &   &   & DART    & 3D & × &   &   &   \\
        \cite{ref:LH18}    & \useurl{code:LH18}    &   & × &   & DRL       &   &   & × & × &   &   &   &   & ODE     & 3D & × &   &   &   \\
        \cite{ref:PA18}    & \useurl{code:PA18}    &   & × &   & DRL       & × &   &   & × &   &   &   &   & Bullet  & 3D & × & × &   & × \\
        \cite{ref:PK18}    & \useurl{code:PK18}    &   & × &   & DRL       &   &   & × & × &   &   &   &   & Bullet  & 3D & × &   &   &   \\
        \cite{ref:YT18}    & \useurl{code:YT18}    &   & × &   & DRL       &   &   &   & × &   &   &   &   & DART    & 3D & × &   &   & × \\
        \cite{ref:AL19}    & \useurl{code:AL19}    &   & × &   & DRL       &   &   & × & × &   &   &   &   & Bullet  & 3D & × & × &   &   \\
        \cite{ref:BC19}    & \useurl{code:BC19}    &   & × &   & DRL       &   &   & × & × &   &   &   &   & Bullet  & 3D & × &   &   &   \\
        \cite{ref:BN19}    & \useurl{code:BN19}    &   & × &   & DRL       &   &   & × & × &   &   &   &   & ODE     & 3D & × & × &   & × \\
        \cite{ref:MA19}    & \useurl{code:MA19}    &   & × &   & DRL       &   &   & × &   & × &   &   &   & MuJoCo  & 3D & × &   &   &   \\
        \cite{ref:MH19}    & \useurl{code:MH19}    &   & × &   & DRL       & × &   &   & × &   &   & × &   & MuJoCo  & 3D & × &   &   &   \\
        \cite{ref:PC19}    & \useurl{code:PC19}    &   & × &   & DRL       &   &   & × & × &   &   &   &   & Bullet  & 3D & × & × &   & × \\
        \cite{ref:PR19}    & \useurl{code:PR19}    &   & × &   & DL \& DRL & × &   & × & × &   &   &   &   & DART    & 3D & × &   &   &   \\
        \cite{ref:RH19}    & \useurl{code:RH19}    &   & × &   & DRL       &   &   &   & × &   &   &   &   & ODE     & 3D & × &   & × & × \\
        \cite{ref:RX19}    & \useurl{code:RX19}    &   & × &   & DRL       & × &   & × & × &   &   &   &   & Unknown & 3D & × &   &   &   \\
        \cite{ref:WL19}    & \useurl{code:WL19}    &   & × &   & DRL       &   &   &   & × &   &   &   &   & DART    & 3D & × & × &   & × \\
        \cite{ref:HB20}    & \useurl{code:HB20}    &   & × &   & DRL       &   &   & × & × &   &   &   &   & Unknown & 3D & × &   &   &   \\
        \cite{ref:KL20}    & \useurl{code:KL20}    &   & × &   & DL        &   &   & × & × &   &   &   &   & Unknown & 3D & × & × & × & × \\
        \cite{ref:MT20}    & \useurl{code:MT20}    &   & × &   & DRL       &   &   & × &   & × &   &   &   & MuJoCo  & 3D & × &   &   &   \\
        \cite{ref:WG20}    & \useurl{code:WG20}    &   & × &   & DL \& DRL & × &   &   & × &   &   &   &   & Bullet  & 3D & × &   &   &   \\
        \cite{ref:XL20}    & \useurl{code:XL20}    &   & × &   & DRL       &   &   &   & × &   &   &   &   & Bullet  & 3D & × & × &   &   \\
        \cite{ref:MY21}    & \useurl{code:MY21}    &   & × &   & DRL       & × &   &   & × &   &   &   &   & Bullet  & 3D & × &   &   &   \\
        \cite{ref:LT06}    & \useurl{code:LT06}    &   &   & × & DL        & × &   &   & × &   &   &   &   & Unknown & 3D & × &   &   &   \\
        \cite{ref:NT15}    & \useurl{code:NT15}    &   &   & × & DL        &   &   & × & × &   &   &   &   & Unknown & 3D & × &   &   &   \\
        \cite{ref:NZ18}    & \useurl{code:NZ18}    &   &   & × & DL        &   &   & × & × &   &   &   &   & Unknown & 3D & × &   &   &   \\
        \cite{ref:JV19}    & \useurl{code:JV19}    &   &   & × & DL \& DRL &   &   & × & × &   &   &   &   & OpenSim & 3D & × &   &   &   \\
        \cite{ref:LP19}    & \useurl{code:LP19}    &   &   & × & DL \& DRL &   &   & × & × &   &   &   &   & OpenSim & 3D & × &   &   &   \\
        \hline
    \end{NiceTabular}
    \label{tab:control}
\end{table*}
    \section{Motion Editing}
\label{section:MotionEditing}
    So far we looked at how motion data can be synthesised, either for predictive or generative purposes, and how to build frameworks for interactively controlling virtual characters. Besides these applications, another important area of character animation is motion editing, which diversifies the creative capabilities of artists. In this section, we focus on \acrShort{DL}-based approaches for motion editing problems divided into three topics: motion cleaning (Section~\ref{subsection:Cleaning}), motion retargeting (Section~\ref{subsection:Retargeting}) and style transfer (Section~\ref{subsection:StyleTransfer}). Finally, we summarise the methods presented in this section in Table~\ref{tab:editing}.

\subsection{Cleaning}
\label{subsection:Cleaning}
    As mentioned by several authors, projection to and inverse projection from learnt manifolds of human motion can be used to clean motion data, \eg to perform operations such as denoising, fixing corrupted information or filling in missing motion sequences. Such problems have for instance been explored using \acrShortPlr{RBM}~\cite{ref:WN15}, convolutional autoencoders~\cite{ref:HS15}, temporal autoencoders~\cite{ref:BB17, ref:LA21}, spatio-temporal \acrShortPlr{RNN}~\cite{ref:WH19}, sequential \acrShortPlr{RNN}~\cite{ref:JL20}, \acrShortPlr{BiLSTM}~\cite{ref:LZ19}, as well as \acrShort{RNN}-based \acrShortPlr{GAN}~\cite{ref:WC21}. While most approaches typically clean motion data through direct projection and inverse projection (\eg~\cite{ref:WN15, ref:HS15}), then fix residual errors as a post-process, it is also possible to include additional constraints during training. For instance, it is possible to enforce bone length constraints and smoothness by including specific loss functions~\cite{ref:LZ19, ref:LZ20}, or to include an additional perceptual loss measuring the difference in high-level features extracted by a pre-trained perceptual autoencoder~\cite{ref:LZ20}, which improves overall visual quality at the cost of a slight increase in reproduction error. Lohit~\etal~\cite{ref:LA21} also proposed to optimise the latent representation to minimise the error between the reconstructed sequence and the correct information of the input sequence, which they applied to filling missing joint trajectories. While most approaches focus on cleaning directly kinematics skeletal data, Holden~\cite{ref:H18} proposed an approach producing joint transforms directly from raw marker data, in a way which is robust to errors in the input data. This approach is based on learning a deep denoising feedforward neural network using marker locations synthetically reconstructed from motion capture data, where the marker data is corrupted in terms of occlusions and positional shifts.
    
    Several authors also proposed to train neural networks to automatically detect ground contact events (\ie whether the feet are in contact with the floor or not), in order to prevent footsliding artefacts. These approaches typically rely on motion data augmented with foot contact information to output foot contact probabilities. To this end, foot contact can be either manually annotated~\cite{ref:SC19}, or automatically computed using different heuristics, usually based on empirical positional~\cite{ref:YK21} or velocity~\cite{ref:ZY20} thresholds. Using such data, different architectures have been proposed to automatically estimate foot contacts, such as relying on a fully connected neural network~\cite{ref:SC19}, on a temporal \acrShort{CNN} with residual connections~\cite{ref:ZY20}, or on an \acrShort{RNN} with \acrShortPlr{GRU} followed by linear layers and ReLU activation~\cite{ref:YK21}. At runtime, the network estimates the ground contact information of each pose, which are then often used within an \acrShort{IK} framework to remove footskate artifacts~\cite{ref:SC19, ref:YK21}. This information can also be included as an additional zero velocity loss within a state-of-the-art method for pose and shape estimation~\cite{ref:ZY20}, or combined with other constraints into a physics-based optimisation~\cite{ref:SG20}. Shi~\etal~\cite{ref:SA20} also identified the importance of foot contacts to mitigate footskating artifacts, and proposed a network predicting simultaneously joint rotations, global root positions, as well as foot contact labels from estimated \DD joint positions, which are then fed into an integrated \acrShort{FK} layer that outputs \DDD positions.

\subsection{Retargeting}
\label{subsection:Retargeting}
    Retargeting~\cite{ref:Gl98} refers to the task of transferring the movements of a source character in a skeletal animation sequence to a target character with a different morphology, \ie to a skeleton with different bone lengths and possibly a different topology. For clarity, and consistently with the literature, in the remainder of this section the source character will be referred to as A and the target character as~B. As pointed out by Aberman~\etal~\cite{ref:AbL20}, there is no formal specification of the task. The purpose at large is to abstract out the dynamics of the source sequence and to reproduce it on a character whose body proportions differ. Effectively, the sought goal is to synthesise a retargeted sequence for the new morphology whose motion mimics the source while remaining visually plausible and natural. Since it is difficult in practice to obtain ground-truth pairs of (source, target) sequences with exactly the same motion, most learning approaches to retargeting, and all the schemes surveyed in this section, rely on unpaired training data without motion correspondences across characters.

    In the seminal work of Villegas~\etal~\cite{ref:VY18}, the retargeting network is built around two \acrShortPlr{RNN}. An encoder \acrShort{RNN} captures the motion context of the source sequence in its hidden state and forwards it to a decoder \acrShort{RNN} that outputs each frame of the retargeted sequence. A reference pose of the target skeleton is provided to the decoder. Besides regularisation loss terms, network training is driven by an adversarial loss and a cycle consistency loss. The adversarial loss attempts to minimise discrepancies between the joint velocities of ground truth (true) and retargeted (fake) sequences. The cycle consistency loss ensures that a motion sequence of character A that is retargeted to B and then back to A remains as close as possible to the original.

    Lim~\etal~\cite{ref:LimC19} and Kim~\etal~\cite{ref:KP20} reported that the above approach tends to generate unrealistic motion. To mitigate this issue, Lim~\etal~\cite{ref:LimC19} retargeted the motion of the root joint separately from the poses at each time step, and combined the results to construct the output sequence. The retargeted poses are computed as joint rotations, represented as quaternions, to be applied to a reference pose of the target skeleton. The cycle consistency loss of Villegas~\etal~\cite{ref:VY18} is replaced by a reconstruction loss associated to self-retargeting to the same character. Kim~\etal~\cite{ref:KP20} argued that \acrShortPlr{CNN} are better suited than \acrShortPlr{RNN} for retargeting because they can more accurately capture the short-term motion dependencies that mostly condition the performance of the task. Accordingly, their scheme relies on a purely convolutional network with temporally dilated convolutions that retargets whole motion sequences in one batch.

    Aberman~\etal~\cite{ref:AbL20} extended the scope of retargeting to skeletons with different topologies, subject to the condition that all considered topologies are homeomorphic. This implies that they can all be reduced to a common \emph{primal skeleton} by merging pairs of adjacent bones. Their representation of skeletal motion based on \emph{armatures} separates the temporally invariant bone offset vectors that define the character morphology from the time-varying bone rotations at each joint that capture the motion dynamics (see Figure~\ref{fig:AbL20}). These two components are processed in distinct parallel branches of the retargeting network. Importantly, a motion sequence is modelled as a graph whose edges correspond to armatures. This provides a principled formalism for processing motion data sampled on the skeletal graph. The retargeting network includes three types of modules: space-time graph-convolutional operators acting on spatio-temporal joint neighbourhoods, graph pooling and graph unpooling operators. Graph pooling merges features of two adjacent edges (armatures) into a single feature. Each graph unpooling operator is designed as the inverse of a graph pooling operator, splitting one edge into two adjacent edges whose features are copied from the original feature. An autoencoder, composed of an encoder followed by a decoder, is learnt for every skeletal structure represented in the training dataset. An encoder fed with motion data for character A generates embeddings of the static bone offset and dynamic joint rotation components for this motion, expressed in the common primal skeleton topology. A decoder for character B maps these embeddings to the retargeted motion for this character. Retargeting is achieved by composing the encoder for the source character with the decoder for the target character.

\begin{table*}
    \setlength{\tabcolsep}{5.9pt}
    \scriptsize
    \centering
    \caption{
        Summary of the methods presented in Section \ref{section:MotionEditing}. Miscellaneous data includes hand-crafted, synthetic, proprietary, unspecified or other public datasets.
    }
    \begin{NiceTabular}{| c | c | *{3}{c} | c | *{7}{c} | *{7}{c} | c |}
        \CodeBefore
            \rowcolors{4}{tablerowcolor1}{tablerowcolor2} % Set alternating row colors
        \Body
        \hline
        & & && & & &&&&&& & &&&&&& & \\
        
        % Headers
        \header{Reference}                          &
        \header{Code}                               &
        \Block{1-3}{\header{Section}} &&            &
        \header{Framework}                          &
        \Block{1-7}{\header{Dataset}} &&&&&&        &
        \Block{1-7}{\header{Architecture}} &&&&&&   &
        \header{Representation}                     \\
        
        & & && & & &&&&&& & &&&&&& & \\
        
        % Sub-headers
        & % no sub-header (Reference)
        & % no sub-header (Code)
        \rotatebox{90}{\subheader{\ref{subsection:Cleaning} Cleaning}}              &
        \rotatebox{90}{\subheader{\ref{subsection:Retargeting} Retargeting}}        &
        \rotatebox{90}{\subheader{\ref{subsection:StyleTransfer} Style Transfer}}   &
        & % no sub-header (Framework)
        \rotatebox{90}{\subheader{CMU \cite{ref:CM03}}}                             &
        \rotatebox{90}{\subheader{Human3.6M \cite{ref:IP14}}}                       &
        \rotatebox{90}{\subheader{HDM05 \cite{ref:MR07}}}                           &
        \rotatebox{90}{\subheader{Holden \etal \cite{ref:HS16}}}                    &
        \rotatebox{90}{\subheader{Mixamo \cite{ref:MI21}}}                          &
        \rotatebox{90}{\subheader{NTU RGB+D \cite{ref:SL16}}}                       &
        \rotatebox{90}{\subheader{Miscellaneous}}                                   &
        \rotatebox{90}{\subheader{Fully Connected}}                                 &
        \rotatebox{90}{\subheader{Convolutional}}                                   &
        \rotatebox{90}{\subheader{Recurrent}}                                       &
        \rotatebox{90}{\subheader{Autoencoder}}                                     &
        \rotatebox{90}{\subheader{RBM}}                                             &
        \rotatebox{90}{\subheader{Variational Autoencoder}}                         &
        \rotatebox{90}{\subheader{Adversarial}}                                     &
        \\ % no sub-header (Representation)
        
        \hline
        \cite{ref:HS15}    & \useurl{code:HS15}    & × &   &   & DL        & × &   &   &   &   &   &   &   & × &   & × &   &   &   &         3D positions        \\
        \cite{ref:WN15}    & \useurl{code:WN15}    & × &   &   & DL        & × &   &   &   &   &   &   &   &   &   & × & × &   &   &          Unknown            \\
        \cite{ref:BB17}    & \useurl{code:BB17}    & × &   &   & DL        & × & × &   &   &   &   &   & × &   &   &   &   &   &   &          $\so{3}$           \\
        \cite{ref:H18}     & \useurl{code:H18}     & × &   &   & DL        & × &   &   &   &   &   & × & × &   &   &   &   &   &   &        Euler Angles         \\
        \cite{ref:LZ19}    & \useurl{code:LZ19}    & × &   &   & DL        & × &   &   &   &   &   &   &   &   & × & × &   &   &   &         3D positions        \\
        \cite{ref:WH19}    & \useurl{code:WH19}    & × &   &   & DL        & × &   & × & × &   &   & × &   &   & × &   &   &   &   &         3D positions        \\
        \cite{ref:JL20}    & \useurl{code:JL20}    & × &   &   & DL        &   & × &   &   &   &   &   &   &   & × & × &   &   & × &          $\so{3}$           \\
        \cite{ref:LZ20}    & \useurl{code:LZ20}    & × &   &   & DL        & × &   &   &   &   &   & × &   &   & × & × &   &   &   &         3D positions        \\
        \cite{ref:SA20}    & \useurl{code:SA20}    & × &   &   & DL        & × & × &   &   &   &   &   &   & × &   &   &   &   & × & 3D positions \& Quaternions \\
        \cite{ref:SG20}    & \useurl{code:SG20}    & × &   &   & DL        &   & × &   &   &   &   & × &   & × &   &   &   &   &   &         2D positions        \\
        \cite{ref:ZY20}    & \useurl{code:ZY20}    & × &   &   & DL        &   & × &   &   &   &   & × &   & × &   &   &   &   &   &         2D positions        \\
        \cite{ref:LA21}    & \useurl{code:LA21}    & × &   &   & DL        &   &   & × &   &   & × &   &   & × &   & × &   &   & × &         2D positions        \\
        \cite{ref:WC21}    & \useurl{code:WC21}    & × &   &   & DL        & × &   &   &   &   &   & × &   &   & × &   &   &   & × &          Unknown            \\
        \cite{ref:YK21}    & \useurl{code:YK21}    & × &   &   & DL        & × &   &   &   &   &   & × &   &   & × &   &   &   &   &           Ad hoc            \\
        \cite{ref:SC19}    & \useurl{code:SC19}    & × &   & × & DL        &   &   &   &   &   &   & × & × &   &   &   &   &   &   &         3D positions        \\
        \cite{ref:VY18}    & \useurl{code:VY18}    &   & × &   & DL        &   & × &   &   & × &   &   &   &   & × &   &   &   & × &         Quaternions         \\
        \cite{ref:LimC19}  & \useurl{code:LimC19}  &   & × &   & DL        &   &   &   &   & × &   &   &   & × &   & × &   &   & × &         Quaternions         \\
        \cite{ref:AbL20}   & \useurl{code:AbL20}   &   & × &   & DL        &   &   &   &   & × &   &   &   & × &   & × &   &   & × & 3D positions \& Quaternions \\
        \cite{ref:KP20}    & \useurl{code:KP20}    &   & × &   & DL        &   & × &   &   & × &   &   &   & × &   &   &   &   & × &         Quaternions         \\
        \cite{ref:HS16}    & \useurl{code:HS16}    &   &   & × & DL        &   &   &   & × &   &   &   &   & × &   & × &   &   &   &         3D positions        \\
        \cite{ref:HoH17}   & \useurl{code:HoH17}   &   &   & × & DL        &   &   &   &   &   &   & × &   & × &   & × &   &   &   &         3D positions        \\
        \cite{ref:WC18}    & \useurl{code:WC18}    &   &   & × & DL        &   &   &   &   &   &   & × &   &   & × & × &   &   & × &          Unknown            \\
        \cite{ref:AW20}    & \useurl{code:AW20}    &   &   & × & DL        &   &   &   &   &   &   & × &   & × &   & × &   &   & × & 3D positions \& Quaternions \\
        \cite{ref:DA20}    & \useurl{code:DA20}    &   &   & × & DL        &   &   &   &   &   &   & × &   & × &   &   &   &   & × &         Quaternions         \\
        \cite{ref:WA20}    & \useurl{code:WA20}    &   &   & × & DL        &   &   &   &   &   &   & × &   &   & × & × &   &   & × &        Euler Angles         \\
        \hline
    \end{NiceTabular}
    \label{tab:editing}
\end{table*}

\subsection{Style Transfer} \label{subsection:StyleTransfer}
    Neural Style Transfer refers to algorithms manipulating data such as images, videos or human animations with \acrShortPlr{DNN} to make the stylistic components look like another data sample. Gatys~\etal~\cite{ref:GE16} first introduced a method to perform neural style transfer on images, using the Gram matrix of the deep features as the artistic style information of an image. In human animation, style transfer aims at transferring the style from one motion sequence to another whose content is retained, called hereafter style and content motion sequences, respectively. For example, we might want to edit a particular motion by affecting the state of mind of the character (\eg enthusiastic, sad, angry) while preserving the performed action, \eg locomotion from point A to point B.

    % unsupervised (Gram matrix)
    In the context of \acrShort{DL}-based skeletal animation, Holden~\etal~\cite{ref:HS16} pioneered human motion style transfer using their deterministic generation model (see Section~\ref{subsection:DeepGenerativeModeling}). In this framework, different types of constraints can be applied on the generated motion, such as trajectory, bone lengths or joint positions, solving a constrained optimisation problem in a learnt human motion manifold. Gradients are back-propagated through an autoencoder representing the manifold to optimise latent representations. Style transfer constitutes a particular case, where both joint positions and style are constrained with respect to the content and style motion sequences, respectively. Moreover, Holden~\etal~\cite{ref:HS16} followed prior work in image style transfer~\cite{ref:GE16} by using the Gram matrix in the latent representation as a style similarity measure and thus do not require style annotations. They further improved this approach by training a feedforward network to perform style transfer thousands of times faster~\cite{ref:HoH17}. Gradients are back-propagated to train the feedforward network instead of optimising the latent representation.
    
    % supervised
    Alternatively to the Gram matrix, style annotations can also be used to guide the learning of a style transfer model. Smith~\etal~\cite{ref:SC19} divided the task of style transfer into spatial and temporal style variations networks, both taking as inputs joint positions as well as a one-hot style vector, where the networks are applied consecutively to predict corresponding style variations. However, this approach requires motion data registered with similar poses in different styles. Wang~\etal~\cite{ref:WC18, ref:WA20} leveraged an \acrShort{LSTM}-based Sequential Adversarial Autoencoder whose encoder learns to map motions to separate content and style encodings: two additional discriminators are trained to recover style labels from content and style encodings, respectively. The encoder tries to fool the former and helps the latter in order to free the content encoding from the style information while preserving it in the style encoding. At runtime, both the content and style motion sequences are encoded into separate content and style embeddings. The decoder combines the content embedding of the content motion sequence and the style embedding of the style motion sequence to produce the style transfer output. Following these works, Aberman~\etal~\cite{ref:AW20} also extracted content and style encodings but with two separate encoders. Furthermore, the style encoder learns a common embedding from both \DD and \DDD joint positions with a triplet loss, which enables style extraction from videos. The output motion is synthesised from the content encoding while the style is controlled by the style encoding through temporally invariant \emph{Adaptive Instance Normalisation} (AdaIN). Moreover, a multi-style discriminator assesses the output motion style.
    
    % Misc
    Beyond general-purpose style transfer, Dong~\etal~\cite{ref:DA20} focused on translation from adult to child motions and vice versa. In this particular case, retargeting is not sufficient since it does not capture the natural stylistic differences between adults and children~\cite{ref:DA20}. This work leverages the capacity of \emph{CycleGAN}~\cite{ref:ZP17} to learn the mapping between adult and child motion distributions without paired training data, which is critical due to the very limited availability of child data.

    \section{Discussion} \label{section:discussion}
	% introduction
	In this state-of-the-art review we explored the recent and promising trends to address challenges in skeleton-based human animation with \acrShort{DL} and \acrShort{DRL}. After a general overview of pose representations and motion data processing with \acrShortPlr{DNN}, we covered motion synthesis, character control and motion editing based on \acrShort{DL} and/or \acrShort{DRL}. In this section, we discuss some of the limitations and potential future work directions.
	
	% representations & modelling
	First, a pose/motion representation should characterise the human motion as precisely as possible, while being suitable for optimisation (\eg avoid the error accumulation problem) and other common applications such as skinning and rigging. However, no simple representation fulfills all these requirements. The success of different approaches~\cite{ref:PG18, ref:PF20, ref:AbL20} suggests that expressing loss functions in the joint positions space is helpful, although the joint orientations are more expressive to represent human poses. To better animate human characters, it is necessary to develop more general pose/motion representation frameworks suitable for learning both temporal and spatial patterns, probably combining the advantages of hierarchical orientations and absolute positions, such as the approach proposed by Aberman~\etal~\cite{ref:AbL20} for skeletal convolutions in the case of motion retargeting (see Figure~\ref{fig:AbL20}). Finally, the spectral domain of human motion remains as of today almost unexplored~\cite{ref:ML19, ref:ML20, ref:CH20} whereas, \eg artifacts related to high-frequencies like noise or footskate might be easier to prevent and/or remove in such a domain.
	
    % motion prediction
	Second, approaches in short-term motion prediction attempt to forecast future motion exactly, with a ground truth being the only solution. Many of them are autoregressive, \ie (partial) outputs are fed back to the model to predict distant future motion frames. As a result, they suffer from instability and tend to either diverge or converge to a mean pose. This is exacerbated when targeting longer horizons since the problem of deterministic prediction becomes more and more ambiguous. Ill-posedness can be mitigated either by adding contextual information or by modeling and sampling from a distribution of possible future outcomes. Hopefully, both short and long term motion prediction could be achieved by a single powerful stochastic model with a quasi-deterministic short-term behaviour, from which stochasticity could emerge in the long term.
    
	% generative models: lack of in-the-wild motion data
	Third, important aspects of generative models in motion synthesis include the quality (\eg perceptual plausibility, absence of artifacts), the intermodal and intramodal diversity, and the level of representation detail (\eg ample arm movements versus subtle finger manipulations), but also the ability to control high-level parameters in synthesised motions. Currently, generative models still fail to fully represent the human motion distribution in the considered scope, with all the diversity and modes it is made of. One reason for this is the lack of in-the-wild motion data, mainly because reliable motion capture systems generally require markers to be placed on the subject captured and a dedicated room for a multi-camera setup. Future advances in marker-less motion capture from image/video might help to collect more and more diverse motion data, and hopefully to build strong human motion generative models, maybe directly from image/video. Moreover, generative models would also benefit from motion data with higher skeleton resolution, \ie from capturing more joints. This would both improve the quality of skinned character animation based on skeletal motion data and enable models to account for more subtle details of human motion, especially those expressed by the hands, the feet and the head. Another promising opening to obtain favorable results with limited amount of in-the-wild data might be self-supervised learning. In such a framework, a representation of the data would first be learnt over in-the-lab data (\ie the pretext task) then fine-tuned over in-the-wild data (\ie the downstream task).
	%
	% generative models: lack of flexibility and interactivity
	Although advances in \acrShort{DL} allow to model the human motion distribution with increasing diversity and fidelity, general-purpose approaches are most of the time too general and lack flexibility. Indeed, the difficulty to embed hard constraints in \acrShort{DL}, \eg desired precise action(s) or physical correctness, and the tendency of neural networks to be black boxes prevent animators from adjusting model behaviours to specific needs or from interactively editing the generated sequences.
	
	% character control
	
	Fourth, unlike with deep generative models used in motion synthesis, interactivity is at the core of character control with models dynamically reacting to user input flows while trying to provide diverse motions that are accurate, physically realistic and performed like real humans would do in a similar situation. On the one hand physics and biomechanics based approaches rely on the laws of physics -- often into multibody physical simulations -- with actuation models, while on the other hand kinematics-based approaches directly produce character motion without hard physical constraints. As of today, the former provide humanoid controllers that are mostly correct physically speaking since hard constraints are part of the models themselves. Although these controllers \eg manage locomotion, jumping and specific actions such as lifting, carrying, or pushing objects, they often fail to imitate human naturalness. It appears to us that the approaches closest to providing natural human motion rely more on real-world data besides their physical model, \eg the framework proposed by Peng~\etal~\cite{ref:PK18} which consists in imitating motions extracted from video using deep pose estimation or the data-driven controller proposed by Bergamin~\etal~\cite{ref:BC19} where the policy network computes corrective offsets added to reference motions. Approaches in kinematics-based character control are exclusively data-driven, which makes it easier to obtain more natural results. However, the lack of physical models/simulation prevents produced animations from being physically correct. In practice, real-world animation pipelines still mostly prefer traditional rather than automated methods due to their unpredictable behaviours and lower quality, although promising advances have been made with \acrShort{DL} and \acrShort{DRL}. Finally, we could expect approaches in character control to increasingly rely on (data-driven) deep generative models together with biomechanical or physical models (\eg~\cite{ref:JV19, ref:LP19, ref:PR19, ref:WG20}) to bring strengths from both types of approaches and converge to diverse, natural, high-quality and physically-plausible controllers. Further advances on the flexibility or the interactivity at runtime are also expected.

    % motion editing
    Fifth, motion editing has still not been explored very extensively even though promising methods have been published (e.g.~\cite{ref:HS15, ref:HS16, ref:AbL20, ref:AW20, ref:DA20}). Yet this topic is important to empower animators in other fields of human animation. A significant contribution in this respect is the motion editing framework proposed by Holden~\etal~\cite{ref:HS16}, where low-level motion parameters learnt by a motion modelling network are mapped to high-level, human understandable controls by means of a disambiguation network. This latter network is trained separately and its inputs can be fine-tuned to the editing task at hand. To the best of our knowledge, this work has not been followed up. In our opinion, the main remaining challenges are the interpretability and controllability of the method, especially in a professional content production workflow. Interpretability refers to determining some relationship between the latent representation and the physical control, while controllability refers to the explicit control of animations given some latent representation. To the best of our knowledge, no method has achieved these two goals at the moment. Removing or preventing artifacts is also a critical part of motion editing, \eg for motion data pre-processing or post-processing in animation workflows. As an example, footskate artifacts are widespread but there is still no successful fully automated method to reliably fix this artifact in motion data. In the future we expect a growing number of works in motion editing at large, and maybe even new topics to emerge like retargeting did 20 years ago.

    % lack of perceptual and user studies in general
    Lastly, most of the studies covered in this survey suffer from a lack of perceptual and user studies, which are crucial. For instance, in motion prediction most works compare their results to others using statistical metrics, such as the mean squared error over predicted joint angles, which are not suited to evaluate how plausible or natural predicted motions would be perceived. Other topics also mostly skip perceptual and user studies although sometimes no ground truth exists, \eg character control. This is particularly striking in motion generation where ground truth is equivocal and the inception score is mainly used to assess the diversity of synthesised motions. We therefore expect more perceptual and user studies as well as performance metrics involving perceptual cues in the future.
    
    % conclusion
    To conclude this survey, we have presented a summary of the advances made over the last few years in human animation based on deep neural networks, either trained using \acrShort{DL} or \acrShort{DRL}. These approaches are skyrocketing today in the field of animation, and will probably increasingly find their place in real-world applications in the entertainment industry to animate synthetic characters.
    \section*{Acknowledgements} \label{section:acknowledgements}
This work was supported by the European Commission under European Horizon 2020 Programme, grant number 951911 - AI4Media.
    \printAcronyms
    \printbibliography                
\end{document}